\documentclass[journal]{IEEEtran}
\usepackage{amsmath,amsfonts, amssymb}
\usepackage{algorithmic}
\usepackage{algorithm}
\usepackage{array}
\usepackage{textcomp}
\usepackage{stfloats}
\usepackage{url}
\usepackage{verbatim}
\usepackage{graphicx}
\usepackage{cite}
\usepackage{multicol}

\usepackage{enumitem}
\setlist{parsep=0pt,listparindent=\parindent}

\usepackage{booktabs,multirow,bm}
\usepackage{tabularx}

\usepackage{xcolor}

\ifCLASSOPTIONcompsoc
  \usepackage[caption=false,font=normalsize,labelfont=sf,textfont=sf]{subfig}
\else
  \usepackage[caption=false,font=footnotesize]{subfig}
\fi

\newcommand{\x}{\mathbf{x}}
\newcommand{\z}{\mathbf{z}}
\newcommand{\y}{\mathbf{y}}

\newcommand{\hz}{\hat{\mathbf{z}}}
\newcommand{\hy}{\hat{\mathbf{y}}}
\newcommand{\hZ}{\hat{Z}}

\newcommand{\bE}{\mathbb{E}}
\newcommand{\bR}{\mathbb{R}}

\newcommand{\bPhi}{\boldsymbol{\phi}}
\newcommand{\bTheta}{\boldsymbol{\theta}}

\DeclareMathOperator*{\argmax}{argmax}

\newtheorem{definition}{Definition} 
\newtheorem{remark}{Remark}

\makeatletter 
\newcommand{\linebreakand}{%
  \end{@IEEEauthorhalign}
  \hfill\mbox{}\par
  \mbox{}\hfill\begin{@IEEEauthorhalign}
}

\makeatother 

\begin{document}
\normalsize
	%
	\title{Robust Information Bottleneck for Task-Oriented Communication with Digital Modulation}
	%
	%
	%
	\author{Songjie~Xie,~\IEEEmembership{Student Member,~IEEE},
                Shuai~Ma,~\IEEEmembership{Member,~IEEE},
                Ming~Ding,~\IEEEmembership{Senior Member,~IEEE}, Yuanming~Shi,~\IEEEmembership{Senior Member,~IEEE}, MingJian~Tang, and Youlong~Wu,~\IEEEmembership{Member,~IEEE} 
             \thanks{The work of Youlong Wu was supported in part by the National Nature Science Foundation of China (NSFC) under Grant 61901267. The work of Yuanming Shi was supported in part by the Natural Science Foundation of Shanghai under Grant No. 21ZR1442700 and Shanghai Rising-Star Program under Grant No. 22QA1406100. The work of Shuai Ma was supported by the National Key Research and Development Program of China (No. 2019YFA0706604) and the Natural Science Foundation (NSF) of China (Nos. 61976169, 62293483). (\emph{Corresponding authors: Youlong Wu and Yuanming Shi.})}
		\thanks{Songjie Xie, Yuanming Shi, and Youlong Wu are with the School of Information Science and Technology, ShanghaiTech University, Shanghai 201210, China (e-mail: \{xiesj, shiym, wuyl1\}@shanghaitech.edu.cn).}
            \thanks{
            Shuai Ma is with the Peng Cheng Laboratory, Shenzhen 518055, China. (e-mail: mash01@pcl.ac.cn).
            }
            \thanks{
            Ming Ding is with the Data61, CSIRO, Sydney, NSW 2015, Australia (e-mail: ming.ding@data61.csiro.au).
            } 
            \thanks{
            M. Tang is with Atlassian, Sydney, NSW 2000, Australia (e-mail: mtang2@atlassian.com).
            }
	}

	\twocolumn
	
	\maketitle
	
\thispagestyle{plain}
\pagestyle{plain}
\setcounter{page}{1}
\begin{abstract}
        Task-oriented communications, mostly using learning-based joint source-channel coding (JSCC), aim to design a communication-efficient edge inference system by transmitting task-relevant information to the receiver. However, only transmitting task-relevant information without introducing any redundancy may cause robustness issues in learning due to the channel variations, and the JSCC which directly maps the source data into continuous channel input symbols poses compatibility issues on existing digital communication systems. In this paper, we address these two issues by first investigating the inherent tradeoff between the informativeness of the encoded representations and the robustness to information distortion in the received representations, and then propose a task-oriented communication scheme with digital modulation, named discrete task-oriented JSCC (DT-JSCC), where the transmitter encodes the features into a discrete representation and transmits it to the receiver with the digital modulation scheme. In the DT-JSCC scheme, we develop a robust encoding framework, named robust information bottleneck (RIB), to improve the communication robustness to the channel variations, and derive a tractable variational upper bound of the RIB objective function using the variational approximation to overcome the computational intractability of mutual information. The experimental results demonstrate that the proposed DT-JSCC achieves better inference performance than the baseline methods with low communication latency, and exhibits robustness to channel variations due to the applied RIB framework. 
	\end{abstract}
	

\section{Introduction}
 \label{sec:intro}
With the rapid development of artificial intelligence (AI) in a variety of domains such as virtual/augmented reality (VR/AR) \cite{anthes2016state, voulodimos2018deep}, brain-computer interfaces \cite{van2012brain}, autonomous vehicular networks \cite{ye2018machine} and smart healthcare \cite{zhou2021review}, the sixth-generation communications (6G) are expected to bring a new paradigm of wireless networks with the assistance of AI \cite{letaief2019roadmap}. 
To support ubiquitous connected intelligent devices and AI services, it is conceived that 6G will beyond the Shannon paradigm to satisfy the stringent requirements including ultra-reliability, low latency, and scalable connectivity \cite{strinati20216g, hoydis2021toward}. Recently, task-oriented communication that extracts and transmits only necessary information for the downstream tasks has received increasing attention from both academia and the industry \cite{shao2021learning, letaief2021edge, shi2023taskoriented}. Different from conventional bit-level communication systems, which ignore the semantic meaning of the transmitted messages, task-oriented communication aims at extracting and transmitting only task-relevant information, which in turn can significantly compress the source, improve communication efficiency and achieve low end-to-end latency. In order to extract and transmit task-relevant information, the learning-based joint source-channel coding (JSCC) \cite{bourtsoulatze2019deep, choi2019neural} technique is widely studied. In learning-based JSCC, deep neural networks (DNNs) are utilized to learn and extract the relevant information and map the extracted task-relevant information to the continuous channel input. It is shown that JSCC outperforms the separate source-channel coding strategy, e.g., in finite block-length coding cases \cite{guyader2001joint} and task-oriented communication systems \cite{shao2021learning}.

Although the learning-based JSCC has achieved empirical success in a variety of areas, two main issues remain in its application in task-oriented communication. The first one is the robustness issue, which comes from the fact that the compacted task-relevant information could be more vulnerable to information distortion. In task-oriented communication, the redundancy elimination reduces the communication overhead, but if no redundancy is added, the information retained in the received representation is more likely to be distorted by perturbation factors such as transmission unreliability, information obfuscation due to data privacy concerns \cite{wei2021low}, etc. The second one is the compatibility issue of JSCC in universal digital communication systems. As the worldwide deployment of 5G \cite{shafi20175g}, modern mobile systems are based on digital modulation \cite{bhagyaveni2016introduction}, while in JSCC the direct transmission of continuous feature representations  needs to be modulated with analog modulation or a full-resolution constellation, which brings huge burdens for resource-constrained transmitter and poses implementation challenges on the current radio frequency (RF) systems. 
In this paper, we aim to address the robustness and compatibility issues described above. Our goal is to investigate the inherent trade-off between the informativeness of the encoded representations and the robustness of communication systems and propose a \emph{discrete} task-oriented JSCC method that jointly extracts task-oriented information to reduce the communication overhead, achieves robustness to channel variations and is compatible with current digital communication systems. 

	\subsection{Related Works}
	\subsubsection{Task-oriented Communication}  The current studies on task-oriented communication facilitate a paradigm shift for communication systems design from the bit-level to the semantic level. 
    There exist certain attempts to apply such a design principle in various tasks.
	In \cite{shao2020bottlenet++}, an effective framework is developed for edge inference with a constraint of on-device computation, where the task-oriented encoding of the intermediate feature is proposed to reduce the communication overhead.
    The same idea is adopted by a generic framework for task-oriented communication \cite{shao2021learning}, which is based on the information bottleneck (IB) principle \cite{tishby2000information}. This framework prunes the redundant dimension of encoded representation to reduce the communication overhead by utilizing the sparsity-inducing variational prior.
	Moreover, the recent work \cite{shao2022task} extended this task-oriented communication framework to multi-device cooperative edge inference by adopting the distributed information bottleneck (DIB) \cite{aguerri2019distributed} for distribute feature encoding. Meanwhile, the IB principle is also adopted in \cite{strinati20216g} to identify the relevant semantic information to accomplish a goal.
	Besides, \cite{xie2021deep} has presented a deep learning based semantic communication system aiming at extracting semantic information from text data and maximizing the semantic transmission rate. A similar strategy is also applied to the image retrieval task \cite{jankowski2020wireless} by transmitting the feature vectors compressed by the DNN-based joint source-channel coding scheme.
    Although the redundancy elimination in task-oriented communication has been proven effective in reducing communication overhead,
    there is a lack of research to investigate the utility of redundancy, particularly for the robustness to information distortion in received representations.
	\subsubsection{Joint Source Channel Coding (JSCC)} The conventional communication system performs source coding and channel coding separately, which guarantees the theoretical optimality in the asymptotic infinite limit of block-length according to Shannon's separation theorem \cite{shannon1948mathematical}. Although the separate design of source coding and channel coding are widely employed in modern communication systems, it is hard to approach the theoretical optimality due to the finite block-length of source and channel code in practice. It is widely perceived that the JSCC owns significant advantages in global optimality compared to separate designs. 
    Over decades, Many JSCC schemes have been proposed in \cite{goldsmith1995joint,guyader2001joint, zhai2005joint, fresia2010joint, chen2018joint}, which are designed to target specific underlying source and channel distribution, and are hard to implement due to the high complexity.
    Inspired by the recent advancement of deep learning, the leaning-based JSCC schemes are proposed for wireless image \cite{bourtsoulatze2019deep, kurka2020deepjscc} and text \cite{farsad2018deep} transmission, which is mainly modeled by DNN-based autoencoders with a noise-injection layer to simulate the noisy channel. Driven by their great performance in data transmission, the learning-based JSCCs have been adopted in inference problems such as wireless image retrieval \cite{jankowski2020wireless} and task-oriented communication systems \cite{shao2021learning, shao2022task}. 

As the transmitted representations are formatted as continuous-valued vectors, the learning-based JSCCs are commonly implemented over analog channels \cite{saidutta2021joint, dai2022nonlinear}. 
 For digital communication, the transmission of continuous values output by JSCC encoders requires the full-resolution constellation design of modulation, which brings a huge burden to the hardware and makes it infeasible to implement in RF systems.
 Although there are certain researches on affordable constellation design for deep learning applications in digital communication \cite{wang2022constellation, xie2020lite, zhu2019joint}, the continuous-valued representation transmission exhibit weak compatibility with digital modulation based wireless systems, which greatly limits the deployment in the current and future communication systems.
 This restriction motivates us to develop a JSCC scheme for task-oriented communication with discrete representations to support digital transmission.
 
	\subsection{Contributions}
	In this work, we propose a theoretical framework for task-oriented communication to enhance the robustness of learned systems to information distortion in received representations.
    Based on the proposed framework, we design a task-oriented communication scheme using discrete representation encoding and digital modulation. The proposed digital task-oriented communication scheme, together with the robust encoding framework, addresses the robustness issue and adapts to digital communication systems. The main contributions of this paper are summarized as follows:
	\begin{itemize}
		\item We propose a robust encoding framework for task-oriented communication, named \emph{Robust Information Bottleneck} (RIB). The proposed RIB framework formulates the informativeness-robustness tradeoff in the encoded representation and aims at maximizing the coded redundancy to improve the robustness, while retaining sufficient information for the downstream inference tasks. Therefore, it addresses the tension between improving the robustness and keeping sufficient relevant information for downstream inference tasks without extra communication overhead.
		\item Due to the computational intractability of mutual information, we derive the tractable variational upper bound of the RIB objective by utilizing the variational approximation technique. The variational distribution is parameterized by the learning-based inference model at the receiver and optimized via end-to-end training. 
		\item Based on the RIB framework, we develop a task-oriented communication scheme with discrete representations to support digital communication, named \emph{Discrete Task-oriented JSCC} (DT-JSCC). Specifically, a probabilistic vector-quantization mechanism is introduced to encode the data samples into the informative discrete representations, and a digital modulation module with a finite-point constellation transmits the discrete representations to the receiver for inference tasks. The proposed DT-JSCC addresses the challenge of the cooperative inference between the transmitter and the receiver under digital transmission. 
		\item To validate the inference performance of the proposed method, we conduct extensive experiments on image classification tasks, which show that the proposed DT-JSCCs outperform the baseline methods and exhibit remarkable robustness to channel variations. Furthermore, the experimental results demonstrate the advantages of the proposed RIB framework in achieving a better informativeness-robustness tradeoff. 
	\end{itemize}
 
	The rest of the paper is organized as follows. The system model is introduced and the problems are described in  Section~\ref{sec:model_des}. In Section~\ref{sec:principle}, we present the robust framework for the learning of task-oriented communication systems. Section~\ref{sec:method} proposes the task-oriented communication schemes based on neural discrete representation learning. The extensive numerical results validate the performance of the proposed task-oriented communication system in Section~\ref{sec:exp}, followed by the discussion and conclusion in Section~\ref{sec:dis} and Section~\ref{sec:con}, respectively.
 
	For notation, we denote random variables by capital letters (e.g. $X$) and their realizations by lowercase letters (e.g. $\x$). We use $\bE[\cdot]$ to denote the statistical expectation. The Shannon entropy of variable $\hat{Z}$ is denoted as $H(\hat{Z})$, the entropy of conditional distribution given $\z$ is denoted as $H(\hat{Z}|\z)$, the mutual information between $\hat{Z}$ and $Y$ is denoted as $I(\hat{Z};Y)$, and the Kullback-Leibler (KL) divergence between two distributions $p(\x)$ and $q(\x)$ is denoted as $D_{KL}(p(\x) \| q(\x))$. We use the subscript to emphasize the dependence of measures on our choice of distribution parameterization, i.e., $H_{\bPhi}(Z|\x) = \bE_{p_{\bPhi}(\z|\x)}[-\log p_{\bPhi}(\z|\x) ]$. Let $[K]$ denote the set $\{1,2, \dots, K\}$ for the positive integer $K \geq 1$. We further denote the circularly symmetric complex Gaussian distribution as $\mathcal{CN}(\cdot, \cdot)$. The notations used in this paper are summarized in Table~\ref{tab:notation}.

 \begin{table}
	\caption{Summary of Notation}
	\label{tab:notation}
	\resizebox{1.0\linewidth}{!}{
	\begin{tabular}{cp{0.65\columnwidth}}
		\toprule
		Notation & Description \\
		\midrule
		$\x, \y$ & Data sample and corresponding target\\
		$\z$ & Encoded representation of data sample $\x$\\
            $d$ & Dimension of encoded representation $\z$, i.e., $\z = (z_1, z_2, \dots, z_d)$ \\
            $K$ & Cardinality of each discrete dimension of $\z$, i.e., $z_j \in [K]$\\
            $h_{\textrm{m}}, g_{\textrm{m}}$ & Modulator and demodulator \\
            $\bPhi, \bTheta$ & The adjustable parameters of the encoder at the transmitter and of the inference model at the receiver.\\
            $f_{\bPhi, j}$ & Feature extractor in the DNN-based encoder \\
            $D$ & Dimension of each extracted feature vector, i.e., $f_{\bPhi, j}(\x) \in \mathbb{R}^D$\\
            $\mathbf{M}$ & Learnable codebook consisting of $K$ $D$-dimensional codewords, i.e., $\mathbf{M}=[\mathbf{m}_1, \mathbf{m}_2, \dots, \mathbf{m}_K]\in \bR^{D\times K}$\\
		\bottomrule
	\end{tabular}}
\end{table}
\section{System Model and Problem Description}
	\label{sec:model_des}
	We consider a point-to-point task-oriented communication system consisting of a single transmitter and a receiver, where the encoded discrete representation is transmitted over the physical channel for the inference tasks. To transmit the discrete representation, the digital modulation scheme is introduced in the system for reliable information transmission. Despite the end-to-end training under a specific channel quality, we aim to learn the discrete representations that are robust to channel quality fluctuation.
	\subsection{System Model}
	\label{subsec:system_model}
	\begin{figure}[t]
		\centering
		\includegraphics[width=8.cm]{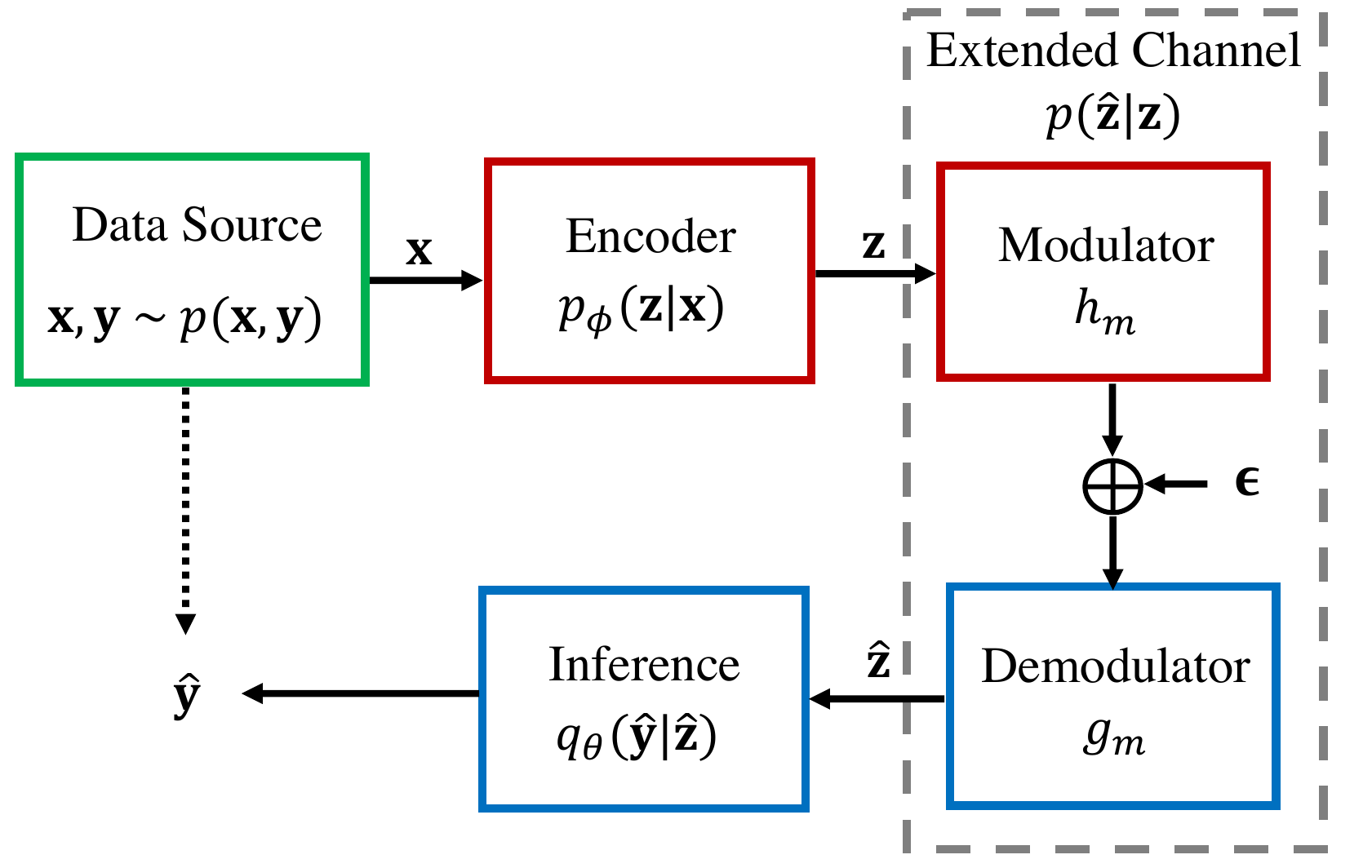}
		\caption{Block diagram of the task-oriented communication system model.
		}
		\label{fig:system-model}
	\end{figure}
 To formalize the task-oriented communication framework, we define a probabilistic system model as shown in Fig.~\ref{fig:system-model}. There is a data source that generates the data $\x$ and the corresponding target $\y$ (e.g. label) with a joint distribution $p(\x,\y)$. Given a dataset $\{\x^{(i)}\}_{i=1}^n$ consisting of $n$ independent and identically distributed (i.i.d.) data points, the communication system aims to estimate the underlying targets $\{\y^{(i)}\}_{i=1}^n$ by conducting cooperative inference between the transmitter and the receiver. 

At the transmitter side, the input data $\x$ is encoded into \emph{discrete} representation vector $\z$ using a probabilistic encoder $p_{\bPhi}(\z|\x)$ parameterized by the adjustable parameters $\bPhi$. The probabilistic model of encoder $p_{\bPhi}(\z|\x)$ is determined by the DNNs deployed on the transmitter.
	We assume that $\z= (z_1, z_2 \dots, z_d) $ is a $d$-dimensional discrete vector and each dimension $z_j \in [K]$ is discrete variable.
	For simplicity, we adopt the mean-field assumption \cite{blei2017variational} and model the $p_{\bPhi}(\z|\x)$ as a factorial categorical distribution with the probability mass function
	\begin{align}
		p_{\bPhi}(\z|\x) &= \prod_{j=1}^d p_{\bPhi}(z_j|\x),
	\end{align}
where the $p_{\bPhi}(z_j|\x)$ is the $K$-way categorical distribution for each dimension $z_j \in [K]$.
	
	Then, the encoded representation $\z$ is modulated into channel input symbols by a digital modulator $h_\textrm{m}$ and transmitted over a noisy channel. Without loss of generality, we consider a general channel model, the additive white Gaussian noise (AWGN) channel, due to its widespread adoption in representing realistic channel conditions. The additive noise vector $\boldsymbol{\epsilon}$ is sampled from an isotropic Gaussian distribution with variance $\sigma^2$, i.e., $\boldsymbol{\epsilon} \sim \mathcal{CN}(\mathbf{0}, \sigma^2\mathbf{I})$. After transmitting over a noisy channel, the noisy symbols are demodulated into the corrupted representation $\hz$ with a demodulation scheme $g_\textrm{m}$.
    With a digital modulation scheme, the modulation, physical channel, and demodulation are considered as an extended channel model from encoded representation $\z$ to corrupted representation $\hz$ with a conditional distribution $p(\hz|\z)$.
	Note that $p(\hz|\z)$ is independent of parameters $\bPhi$ of the encoder at the transmitter, we obtain a Markov chain of the above variables:
	\begin{align}
		Y&\longleftrightarrow X \stackrel{\bPhi}{\longleftrightarrow} Z \longleftrightarrow  \hZ .\label{eq:markov_chain}
	\end{align}
	From the encoding perspective, the encoder $p_{\bPhi}(\z|\x)$ induces the conditional distribution $p_{\bPhi}(\hz|\y)$, the marginal distribution $p_{\bPhi}(\hz)$, and the corresponding posterior $p_{\bPhi}(\y|\hz)$ by the following integrals and application of Bayes law:
	\begin{align}
		p_{\bPhi}(\hz) & = \int p(\x, \y) p_{\bPhi}(\hz|\y) d\x d\y,    \\
		p_{\bPhi}(\y| \hz) & = \frac{\int p(\x, \y)p_{\bPhi}(\hz|\y)d\x}{p_{\bPhi}(\hz)}. \label{eq:intract_posterior}
	\end{align}
	
	Theoretically, the optimal inference model of the encoder $p_{\bPhi}(\z|\x)$ and the channel model $p(\hz|\z)$ is given by the posterior $p_{\bPhi}(\y|\hz)$. However, due to the intractability of the high-dimensional integrals in the posterior computation, we replace the optimal inference model with a variational approximation $q_{\bTheta}(\y|\hz)$ parameterized by adjustable parameters $\bTheta$. At the receiver side, the inference model $q_{\bTheta}(\y|\hz)$ receives the corrupted representation $\hz$ and performs inference to output the result $\hy$.   
 
	Note that the considered task-oriented communication model includes the JSCC scheme and the modulation module. One can extend this probabilistic model to the JSCC scheme with continuous representation $\z$ by substituting digital modulation with analog modulation.

	\subsection{Problem Description} \label{subsec:des}
 	\begin{figure*}[t]
		\centering
		\subfloat[]{
			\centering
			\includegraphics[width=0.2\linewidth]{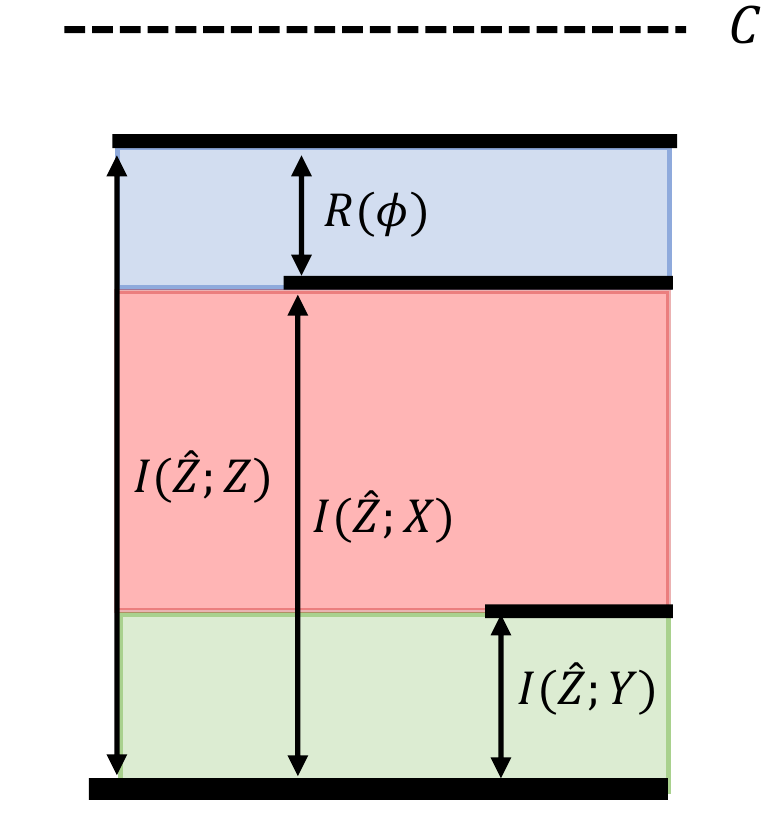}
              \label{subfig:illu_tradeoff_0}
		}
		\subfloat[]{
			\centering
			\includegraphics[width=0.20\linewidth]{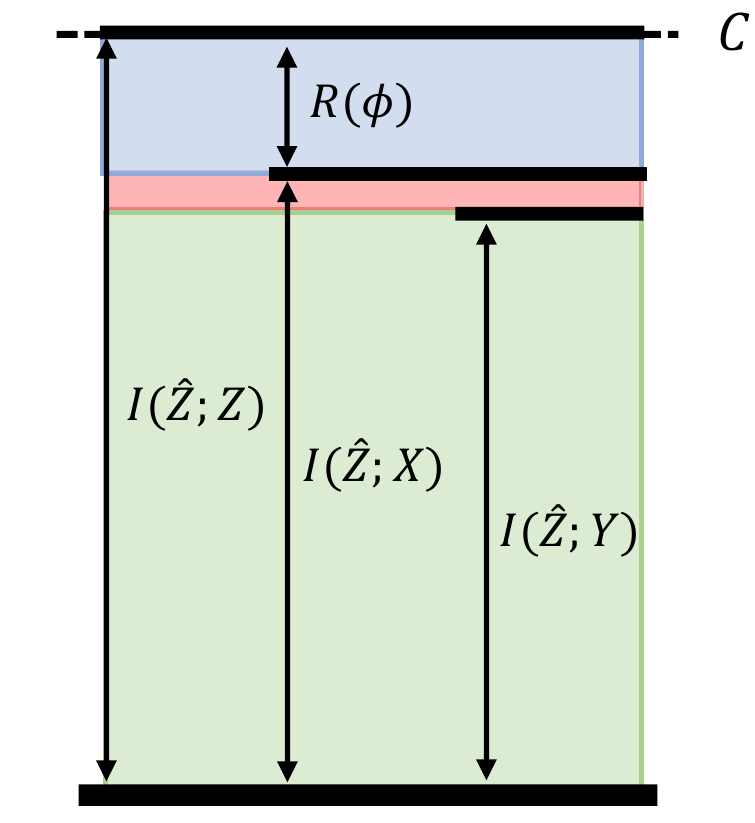}
              \label{subfig:illu_tradeoff_1}
		}
		\subfloat[]{
			\centering
			\includegraphics[width=0.20\linewidth]{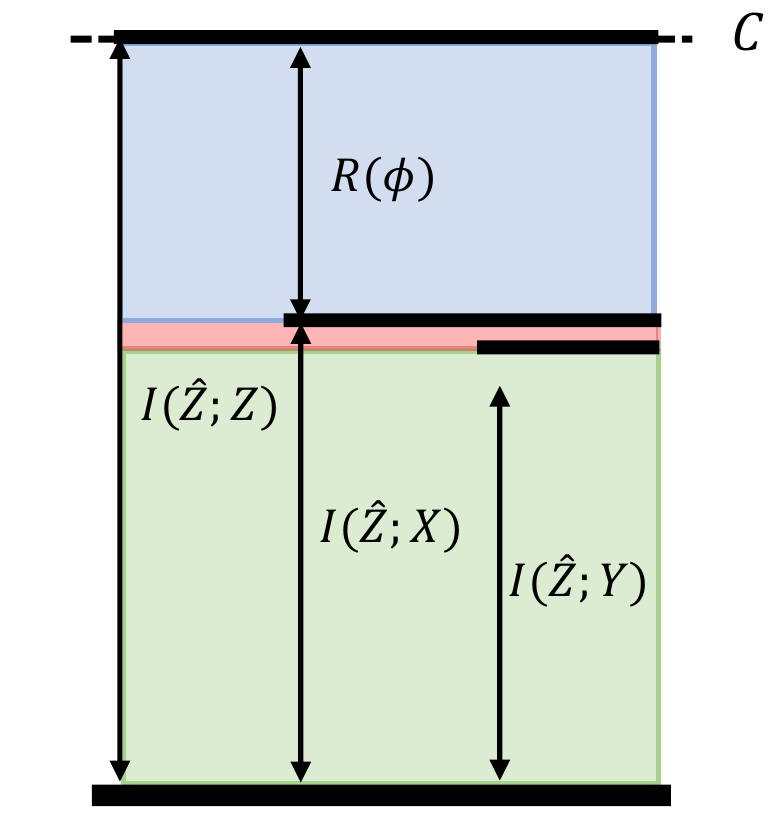}
              \label{subfig:illu_tradeoff_2}
		}
		\subfloat[]{
			\centering
			\includegraphics[width=0.19\linewidth]{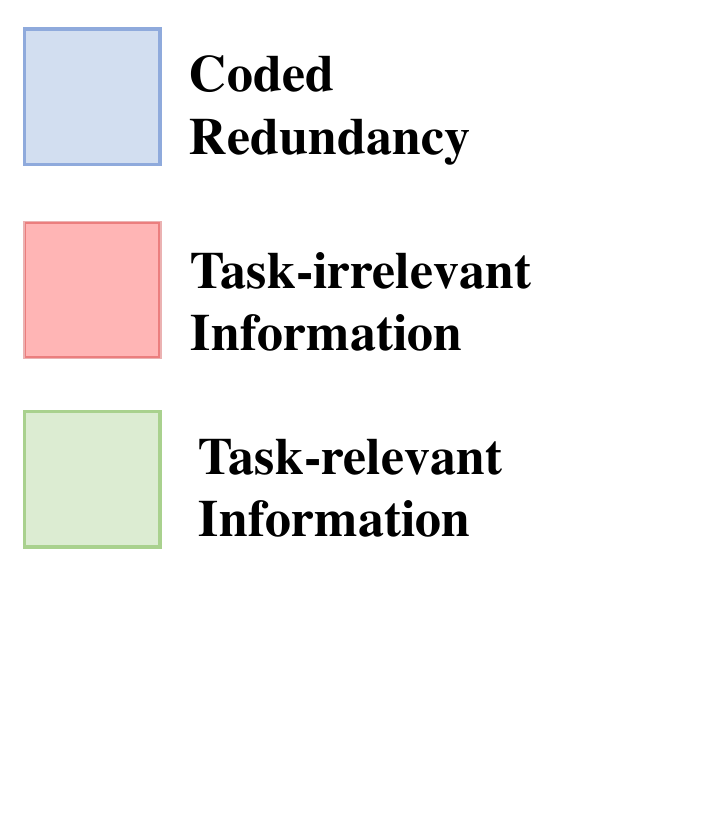}
		}
		\caption{(a): The illustration of the relation between the channel capacity $C$, the transmission rate $I(\hZ; Z)$, coded redundancy $R(\bPhi)$, the amount of task-relevant information $I(\hZ;Y)$ and task-irrelevant information  $I(\hZ;X)-I(\hZ;Y)$. 
			(b): The RIB with a small value of $\beta$ induces that maximal transmission rate $I(\hZ; Z)=C$, a large amount of task-relevant information, and the small coded redundancy. 
			(c): The RIB with a big value of $\beta$ induces that maximal transmission rate $I(\hZ; Z)=C$, the small amount of task-relevant information, and the large redundancy encoded in the representation $Z$ to improve the robustness of the communication system.
		}
		\label{fig:illu_tradeoff}
	\end{figure*}
To design a communication-efficient model for downstream inference tasks, the encoder $p_{\bPhi}(\z|\x)$ needs to extract the informative messages about the target $Y$, while neglecting the irrelevant information in $Z$ given $X$. From the view of data compression, the optimal $Z$ is the \emph{minimal sufficient statistics} \cite{kullback1951information} of $X$ for target $Y$, and can be approximated through the optimization of the information bottleneck (IB) \cite{tishby2000information} problem where the $I(Y;\hat{Z})$ is maximized with the constraint on the amount of the preserved information $I(X; \hat{Z})$, computed by
\begin{align}
			\max\limits_{p_{\bPhi}(\z|\x)}  I(Y;\hZ) -\beta I(X; \hZ). \label{eq:IB}
\end{align}
Apart from data compression, another essential goal of communication system design is to maximize the transmission rate, which is attained by maximizing the mutual information between the transmitted representation $\z$ and the received representation $\hz$:
\begin{align}
    \max\limits_{p_{\bPhi}(\z)}\ I(Z;\hat{Z}), \label{eq:channel_coding}
\end{align}
where $p_{\bPhi}(\z)$ is the marginal distribution depending on the parameters $\bPhi$.

We aim to maximize the relevant information in $Z$ about target $Y$ considering the stochastic process of corrupting $Z$ into $\hat{Z}$. 
By combining the goals of relevant information extraction and transmission rate maximization,  we propose a new principle, Robust Information Bottleneck (RIB), which is formulated by an optimization problem that maximizes the following objective function
	\begin{equation}
		\max\limits_{p_{\bPhi}(\z|\x)} I(Y;\hZ) +\beta[\underbrace{I(Z; \hZ) - I(X;\hZ)}_{R(\bPhi)}],
		\label{eq:RIB}
	\end{equation}
for some fixed $\beta \geq 0$. To interpret the RIB principle, we further investigate the encoding process from the target $Y$ to the received representation $\hat{Z}$ in task-oriented communication by defining the following quantities. 
\begin{definition}[Task-relevant information]
Task-relevant information is the preserved information in $\hat{Z}$ about target $Y$, which is characterized by $I(Y;\hat{Z})$.
\end{definition}
\begin{definition}[Task-irrelevant information]
The amount of task-irrelevant information is measured by $I(X; \hat{Z}) - I(Y; \hat{Z})$, where $I(X; \hat{Z})$ is the amount of total information encoded in $\hat{Z}$ given $X$.  
\end{definition}
\begin{definition}[Coded redundancy]
Coded redundancy $R(\bPhi)$ refers to the amount of redundancy deliberately introduced in the reconstructed representation $\hat{Z}$, defined by $R(\bPhi) \triangleq I(Z; \hat{Z}) - I(X; \hat{Z})$, which is utilized to improve the robustness of the communication system against information distortion that may occur in the received representation $\hat{Z}$. \label{def:coded_redundancy}
\end{definition}

From the underlying Markov chain presented in (\ref{eq:markov_chain}), the relation among the above quantities can be obtained by the data processing inequality \cite{cover1999elements}:
 \begin{equation}\label{eq:key-DPI}
		I(Y;\hZ) \leq I(X; \hZ) \leq I(Z; \hZ) \leq C, 
\end{equation}
where $C$ denotes Shannon's channel capacity of extended channel $p(\hz|\z)$, defined by $C = \max_{p_{\bPhi}(\z)}I(Z; \hZ)$.
As illustrated in Fig.~\ref{subfig:illu_tradeoff_0}, the transmission rate is the summation of task-oriented information, task-irrelevant information, and coded redundancy. Note that $I(Z;\hat{Z})$ is upper bounded by the channel capacity $C$, there is a tradeoff between the task-relevant information and the coded redundancy when the task-irrelevant information is minimized in representation $\hat{Z}$.

Therefore, as presented in (\ref{eq:RIB}), the optimization problem of the proposed RIB principle formulates this inherent informativeness-robustness tradeoff controlled by the parameters $\beta$. Specifically, a small value of $\beta$ privileges inference performance in a specific channel condition by retaining maximal task-relevant information, whereas a large value of $\beta$ privileges the robustness against information distortion in $\hat{Z}$ by preserving more redundancy in $\hat{Z}$, as illustrated in Fig.~\ref{subfig:illu_tradeoff_1} and Fig.~\ref{subfig:illu_tradeoff_2} respectively. 
Furthermore, regardless of the value of $\beta$, the maximization of the RIB optimization problem is equivalent to maximizing the transmission rate $I(Z;\hZ)$ and minimizing the amount of task-irrelevant information simultaneously. In summary, the key idea of the RIB principle is to \emph{keep minimal but sufficient task-relevant information and leave the rest redundancy utilized for robust encoding.}
\section{Robust Encoding for Task-oriented Communication}
In this section, we develop a robust encoding framework based on the proposed RIB principle. Furthermore, we derive a variational upper bound of the RIB objective function by leveraging a variational distribution parameterized by the DNN on the receiver. 
	\label{sec:principle}
	\subsection{Robust Encoding with Robust Information Bottleneck}
	We now turn our attention to how the RIB optimization problem is solved to develop robust task-oriented communication systems. To develop a robust encoder $p_{\bPhi}(\z|\x)$, we turn the RIB optimization problem presented in (\ref{eq:RIB}) into the objective function $\mathcal{L}_{\text{RIB}}(\bPhi)$:
	\begin{align}
		&\min\limits_{p_{\bPhi}(\z|\x)}-I(Y;\hZ) -\beta[I(Z; \hZ) - I(X;\hZ)]  
		\\
		=&\min\limits_{\bPhi}\,\bE_{p(\x,\y)}\{\bE_{p_{\bPhi}(\hz|\x)}[-\log p_{\bPhi}(\y|\hz)] \notag \\ 
		&\qquad  + \beta\bE_{p_{\bPhi}(\z|\x)}[H(\hZ|\z)]- \beta H_{\bPhi}(\hZ|\x)\}- H(Y)   \\
		\overset{\textrm{(a)}}{=}&\min\limits_{\bPhi}\,\bE_{p(\x,\y)}\big\{\underbrace{\bE_{p_{\bPhi}(\hz|\x)}[-\log p_{\bPhi}(\y|\hz)]}_{\textrm{Distortion}} \notag \\ 
		&\qquad  + \underbrace{\beta \bE_{p_{\bPhi}(\z|\x)}[H(\hZ|\z)]- \beta H_{\bPhi}(\hZ|\x)}_{-\beta R(\bPhi)}\big\} \label{eq:objective}\\
		\triangleq &\min\limits_{\bPhi}\;\mathcal{L}_{\textrm{RIB}}(\bPhi), 
	\end{align} 
	where the equivalence (a) holds by ignoring the constant terms $H(Y)$ given the data source. 
 In \eqref{eq:objective}, the first term represents the accuracy of the inference model with the corrupted representation $\hz$, denoted as the \emph{distortion} term. The subsequent terms, denoted by $-\beta R(\bPhi)$, aim to maximize the coded redundancy and enhance the robustness of the system. The second term maximizes the transmission rate for a given channel model $p(\hz|\z)$, while the third term measures the uncertainty of $\hat{Z}$ given data $\x$. By jointly maximizing the transmission rate and uncertainty of $\hZ$, the system retains more redundancy in $\hat{\z}$ to handle information distortion caused by channel noise, ultimately improving the robustness of the task-oriented communication system. The parameter $\beta \geq 0$ controls the tradeoff between inference performance and model robustness.
 \subsection{Variational Upper Bound of Robust Information Bottleneck Objective}
 \label{subsec:variational}
 In the distortion term of objective $\mathcal{L}_{\text{RIB}}(\bPhi)$, the computation of the posterior $p_{\bPhi}(\y|\hz)$ is intractable for the most high-dimensional data sources, as we state in \eqref{eq:intract_posterior}.
	We resort to the variational Bayesian method to approximate $p_{\bPhi}(\y|\hz)$, which has been widely used in machine learning applications such as variational autoencoder \cite{kingma2013auto} and variational information bottleneck \cite{alemi2016deep}. 
	Specifically, we adopt the inference model $q_{\bTheta}(\y|\hz)$ as a variational approximation where $\bTheta$ denotes learnable the parameters of the neural networks at the receiver. By introducing the variational distribution $q_{\bTheta}(\y|\hz)$, we derive the variational upper bound of the distortion term in (\ref{eq:objective}) as follows: 
	\begin{align}
		&\bE_{p(\x,\y)}\{\bE_{p_{\bPhi}(\hz|\x)}[-\log p_{\bPhi}(\y|\hz)]\} \notag\\
		=& \bE_{p(\x,\y)}\{\bE_{p_{\bPhi}(\hz|\x)}[-\log q_{\bTheta}(\y|\hz)]\} \notag\\
		&\qquad \qquad \quad - \underbrace{ \bE_{p_{\bPhi}(\hz)}\{\bE_{p_{\bPhi}(\y|\hz)}[\log \frac{p_{\bPhi}(\y|\hz)}{q_{\bTheta}(\y|\hz)}]\}}_{D_{\text{KL}}(p_{\bPhi}(\y|\hz)||(q_{\bTheta}(\y|\hz))} \\
		\leq& \bE_{p(\x,\y)}\{\bE_{p_{\bPhi}(\hz|\x)}[-\log q_{\bTheta}(\y|\hz)]\}, 
	\end{align}
	where the last inequality follows from the non-negative KL-divergence $D_{\text{KL}}(p_{\bPhi}(\y|\hz)||(q_{\bTheta}(\y|\hz)) \geq 0$. Furthermore, since $\hZ$ is the transmitted representation corrupted by additional Gaussian noise, we can derive a lower bound for the entropy term $H_{\bPhi}(\hZ|\x)$ as follows: 
 \begin{align}
H_{\bPhi}(\hZ|\x) \geq H_{\bPhi}(Z|\x).
\end{align}
 
 Having proposed the above variational upper bound of the distortion term, the tractable variational objective function $\mathcal{L}_{\text{VRIB}}(\bPhi,\bTheta)$ is obtained by:
	\begin{align}
		\mathcal{L}_{\text{RIB}}(\bPhi) &\leq \mathcal{L}_{\text{VRIB}}(\bPhi, \bTheta)  \\
		& \triangleq \bE_{p(\x,\y)}\big\{\bE_{p_{\bPhi}(\hz|\x)}[-\log q_{\bTheta}(\y|\hz)]\notag \\
		&\qquad \quad + \bE_{p_{\bPhi}(\z|\x)}[H(\hZ|\z)]-\beta H_{\bPhi}(Z|\x)\big\}.  \label{eq:variational_objective}
	\end{align}
	Recalling the probabilistic model of the task-oriented communication, the DNN-based encoder is defined by the factorial distribution $p_{\bPhi}(\z|\x) = \prod_{i=1}^d p_{\bPhi}(z_i|\x)$, and the channel model is given by the conditional distribution $p(\hz|\z)$. The conditional entropy terms in $\mathcal{L}_{\textrm{VRIB}}(\bPhi,\bTheta)$, $H(\hZ|\z)$ and $H_{\bPhi}(Z|\x)$, are analytically computable with respect to the parameters $\bPhi$.  Specifically, the entropy terms can be decomposed into the following summations:
 \begin{align}
		H_{\bPhi}(Z|\x) & = \sum\limits_{j=1}^d H_{\bPhi}(Z_j|\x), \label{eq:entropy_x}\\
		H(\hZ|\z) & = \sum\limits_{j=1}^d H(\hat{Z}_j|z_j).\label{eq:entropy_z}
\end{align}
	Given a dataset of $N$ data points $\{\x^{(i)}\}_{i=1}^{N}$ as well as the corresponding labels $\{\y^{(i)}\}_{i=1}^N$, we can now form the Monte Carlo estimation for $\mathcal{L}_{\text{VRIB}}(\bPhi, \bTheta)$ by sampling $L$ realizations of $\z$ from the encoder $p(\z|\x)$ with physical channel noise $\boldsymbol{\epsilon}$, the deterministic modulator $h_{\text{m}}$ and the deterministic demodulator $g_{\text{m}}$. We have the Monte Carlo estimate $\tilde{\mathcal{L}}_{\text{VRIB}}(\bPhi, \bTheta) \simeq \mathcal{L}_{\text{VRIB}}(\bPhi, \bTheta)$:
	\begin{align}
		&\tilde{\mathcal{L}}_{\text{VRIB}}(\bPhi, \bTheta)  = \frac{1}{N}\sum\limits_{i=1}^N\{\frac{1}{L}\sum\limits_{l=1}^L[-\log q_{\bTheta}(\y^{(i)}|\hz^{(i,l)}) \notag \\
		&\qquad \quad + \beta \sum\limits_{j=1}^d H(\hat{Z}_j|z_j^{(i,l)}) ] -\beta \sum\limits_{j=1}^d H_{\bPhi}(Z_j|\x^{(i)})\},\label{eq:RIB_MC}\\
		&\text{where  }\ \hz^{(i,l)} = (\hat{z}^{(i,l)}_j)_{j=1}^d,\ \hat{z}_j^{(i,l)} = g_{\text{m}}(h_{\text{m}}(z_j^{(i,l)})+\epsilon_j^{(i,l)}), \notag\\
		&z_j^{(i,l)} \sim p_{\bPhi}(z_j|\x^{(i)}), \ \text{and}\ \epsilon_j^{(i,l)} \sim \mathcal{CN}(0, \sigma^2) \notag .
	\end{align}
   \begin{figure*}[t]
		\centering
		\includegraphics[width=0.8\textwidth]{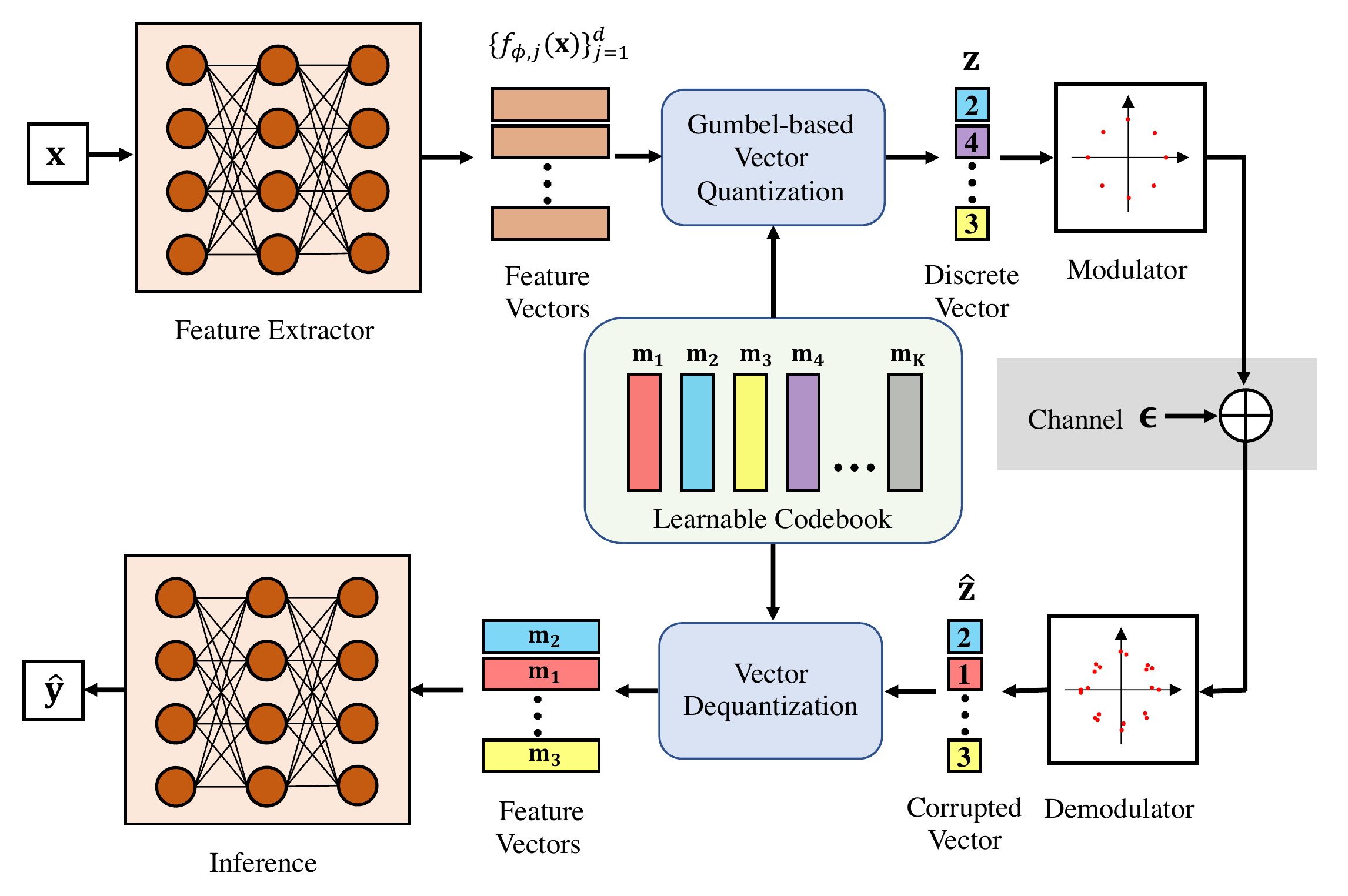}
		\caption{The proposed task-oriented communication scheme with discrete representations, DT-JSCC, where the feature extractor, the learnable codebook, and the inference model are optimized jointly via end-to-end training. }
		\label{fig:DT-JSCC}
\end{figure*}
 \begin{remark}\label{rem:1}
 In the proposed robust encoding framework, RIB, the variational optimization is developed without introducing any prior distribution on the intractable marginal distribution $p_{\bPhi}(\z)$. Therefore, the optimization by the RIB framework does not restrict the learned marginal distribution $p_{\bPhi}(\z)$ to be close to a given prior, allowing it to maximize the transmission rate and reach the capacity of the extended channel (see Fig. \ref{fig:ablation-channelCapacity} in the later experiments section). This flexibility in optimizing the transmission rate without being constrained by prior assumptions is a significant advantage of the RIB framework over the variational IB frameworks \cite{alemi2016deep, shao2021learning}.
\end{remark}
 \begin{remark}
     The robust encoding framework instantiates the RIB principle and provides a generic design for robust task-oriented communication.
     It can be applied not only to task-oriented communication systems with continuous $\z$ but also to systems with discrete $\z$, which makes it possible to further propose a task-oriented communication system optimized jointly with digital modulation. 
 \end{remark}

\section{Task-oriented Joint Source Channel Coding with Digital Modulation}
\label{sec:method}
	The RIB-based framework developed in Section~\ref{sec:principle} is a generic framework for robust encoding. Considering the challenges of continuous representation transmission in digital communication, we propose to transmit the discrete representations in cooperative inference between the transmitter and the receiver.
    In this section, we design a robust task-oriented communication scheme with digital modulation, named discrete task-oriented JSCC (DT-JSCC), where the task-relevant information and coded redundancy are encoded into a vector of integers, and transmitted for downstream tasks by digital modulation.
	
	An overview of our proposed framework is shown in Fig.~\ref{fig:DT-JSCC}. We first encode the data sample $\x$ into the discrete representation vector $\z$ by the DNN-based encoder and the trainable codebook stored at both the transmitter and the receiver. Then, the discrete vector $\z$ is modulated into the channel input symbols, and those symbols are transmitted over bandwidth-limited wireless channels. On the receiver, the noisy transmitted symbols are demodulated into the corrupted discrete representation vector $\hz$, and the inference model leverages $\hz$ with the trainable codebook to output the inference result.
	
	\subsection{DT-JSCC}
        The inference procedure in the proposed task-oriented communication scheme, DT-JSCC, can be summarized into three steps: \emph{feature splitting}, \emph{Gumbel-based vector quantization} and \emph{digital modulation}. 
        \begin{figure}
		\centering
		\subfloat[Feature splitting for dense layers]{
			\centering
			\includegraphics[width=0.85\linewidth]{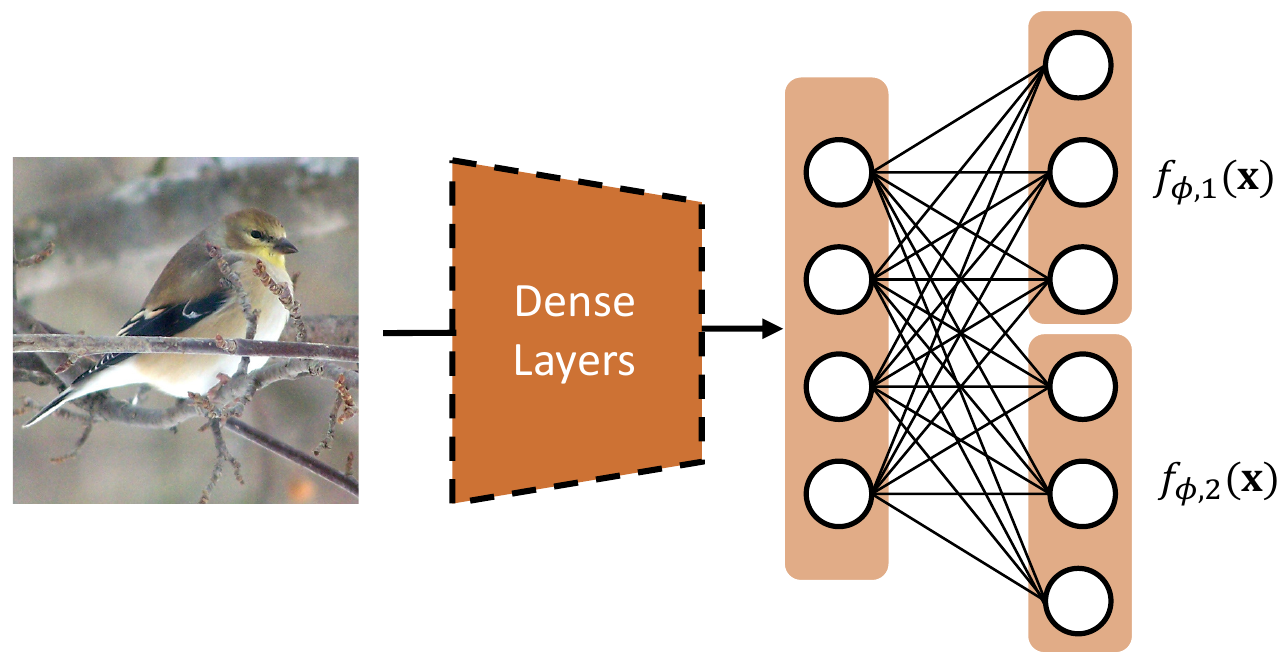}\label{subfig:fully-conn}
		}
  
		\subfloat[Feature splitting for convolutional layers]{
			\centering
			\includegraphics[width=0.85\linewidth]{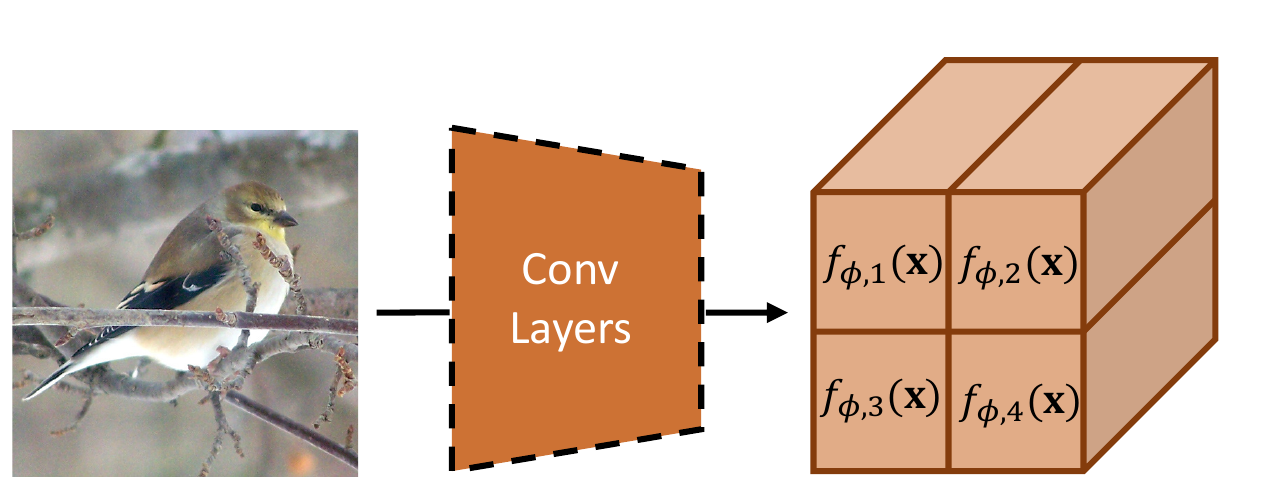}\label{subfig:conv}
		}
		\caption{The proposed strategy to split the encoded features into $d$ vectors $\{f_{\phi, j}(\x)\}_{j=1}^d$ for the neural network with (a) dense layers and (b) convolutional layers.}
		\label{fig:feature-splitting}
	\end{figure}
	\subsubsection{Feature Splitting}
        Since the intermediate features produced by a DNN-based feature extractor are high-dimensional, the vector quantization on the entire features requires a large codebook to guarantee the quantization quality, which leads to heavy memory occupation at both the transmitter and the receiver. In order to quantize the intermediate features with a small-scale codebook,  we propose to equally split the high-dimensional intermediate features into $d$ feature vectors with lower dimensionality. Formally, we define $d$ DNN-based feature extractor functions $\{f_{\bPhi, j}\}_{j=1}^d$ where $f_{\bPhi,j}:\mathcal{X} \to \mathbb{R}^{D}$ to output $d$ feature vectors $\{f_{\phi, j}(\x)\}_{j=1}^d$.
        Specifically, we adopt different feature splitting strategies for the DNN with different structure of layers. For the neural networks with dense layers (i.e., fully connected layers), we equally split the output vector of the last dense layer into a set of feature vectors as illustrated in Fig.~\ref{subfig:fully-conn}. For the neural networks with convolutional layers, the features produced by the last convolutional layer are simply split among channels as illustrated in Fig.~\ref{subfig:conv}.

	\subsubsection{Gumbel-based Vector Quantization}\label{subsubsec:gumbel}
	As shown in Fig.~\ref{fig:DT-JSCC}, the $d$ extracted feature vectors $\{f_{\bPhi, j}(\x)\}_{j=1}^d$ are vector-quantized into a $d$-dimensional discrete vector $\z = (z_1, z_2, \dots, z_d)\in [K]^d$. 
 
 We start by defining a learnable codebook, denoted by $\mathbf{M}=[\mathbf{m}_1, \mathbf{m}_2, \dots, \mathbf{m}_K]\in \bR^{D\times K}$, which consists of $K$ $D$-dimensional codewords. This codebook is a set of parameters shared between the encoder on the transmitter and the inference network on the receiver, and is part of the overall parameter set $\bPhi$ and $\bTheta$ that are trained jointly in an end-to-end manner through backpropagation.

 To obtain the associated logits for a given feature vector $f_{\bPhi, j}(\x)$, we directly project it onto the codebook $\mathbf{M}$. Specifically, the logits $\boldsymbol{\ell}_j$ are computed as the inner product between $\mathbf{M}$ and $f_{\bPhi, j}(\x)$,
  \begin{align}
            \boldsymbol{\ell}_j &=  \mathbf{M}^T f_{\bPhi,j}(\x) \label{eq:logits}\\
            &= (\mathbf{m}_1^Tf_{\bPhi,j}(\x), \mathbf{m}_2^Tf_{\bPhi,j}(\x),\dots, \mathbf{m}_K^Tf_{\bPhi,j}(\x))\\
		&= (\ell_{j,1}, \ell_{j,2},\dots, \ell_{j, K}) ,
\end{align}
where each $\ell_{j,k}$ of the resulting logits $\boldsymbol{\ell}_j$ represents the similarity score between $f_{\bPhi,j}(\x)$ and the $k$-th codeword $\mathbf{m}_k$.

Given the logits $\boldsymbol{\ell}_j$, the probabilistic encoder $p_{\bPhi}(z_j|\x)$ is modeled as a categorical distribution. Specifically, the probability of index $k$ in the discrete representation vector $\z$ is determined by the softmax function of the corresponding logit $\ell_{j,k}$,
\begin{align}
		p_{\bPhi}(z_j=k|\x) &= \frac{\exp(\ell_{j,k})}{\sum\limits_{k'=1}^{K} \exp(\ell_{j,k'})}. \label{eq:posterior}
	\end{align}
The above conditional distribution $p_{\bPhi}(z_j|\x)$ is used to compute the entropy term $H_{\bPhi}(Z|\x)$ of $\mathcal{L}_{\text{VRIB}}(\bPhi, \bTheta)$, as presented in \eqref{eq:entropy_x}.

    In order to perform Monte Carlo estimation of $\mathcal{L}_{\text{VRIB}}(\bPhi, \bTheta)$, we need to sample from the categorical distribution $p_{\bPhi}(z_j|\x)$. However, direct sampling is not differentiable, so we use the Gumbel-softmax distribution and its continuous reparameterization technique, known as the \emph{Gumbel-softmax trick}\cite{jang2016categorical}, to make the probabilistic model differentiable and trainable with backpropagation.

    The Gumbel-Softmax trick is a technique that enables differentiable sampling from a categorical distribution, commonly used to train neural networks with discrete output spaces. This is achieved by adding Gumbel noise to the logits and applying the softmax function, allowing for continuous optimization of categorical variables through temperature annealing.
    Specifically, to obtain a differentiable approximation of categorical sampling for $z_j$ using logits $\boldsymbol{\ell}_j$, we compute a sample vector $\mathbf{w}_j= (w_{j,1},w_{j,2}, \dots, w_{j,K}) \in \mathbb{R}^K$:
	\begin{align}
		w_{j,k} &= \frac{\exp((\ell_{j,k}+\upsilon_k)/\tau)}{\sum\limits_{k'=1}^{K} \exp((\ell_{j,k'}+\upsilon_{k'})/\tau)}, \label{eq:sample_vec}
	\end{align}
	where $\upsilon_1, \upsilon_2, \dots, \upsilon_K$ are i.i.d samples from the Gumbel distribution computing by $\upsilon_k= -\log(-\log(u_k))$ where $u_k\sim \text{Uniform}(0,1)$, and $\tau \geq 0$ is the softmax temperature. As $\tau$ approaches $0$, the sample vector $\mathbf{w}_j$ is close to the one-hot like vector and the Gumbel-softmax distribution approaches the categorical distribution $p_{\bPhi}(z_j|\x)$. With the Gumbel-Softmax trick, the sampling of categorical distribution $p_{\bPhi}(z_j|x)$ is estimated by using argmax function on the vector $\mathbf{w}_j$:
 \begin{align}
     z_j &= \argmax_{k\in [K]} w_{j,k}.\label{eq:gumbel_max}
 \end{align}

	On the transmitter, we compute the sample vector $\mathbf{w}_j$ with (\ref{eq:sample_vec}) for each dimension $z_j$. Then, the discrete representation vector $\z$ is generated by sampling from $p_{\bPhi}(\z|x)$ based on (\ref{eq:gumbel_max}) for each discrete dimension $z_j$, and backpropagate the differentiable samples vectors $\{\mathbf{w}_j\}_{j=1}^d$ with an annealing temperature $\tau$.
        For the dequantization part at the receiver, the corrupted discrete vector $\hz$ is mapped into a set of feature vectors $\{\mathbf{m}_{\hat{z}_j}\}_{j=1}^d$ with the learnable codebook $\mathbf{M}$.

\begin{algorithm}[t]
\small
\caption{Training DT-JSCC}
\begin{algorithmic}[1]
\label{algo-DTJSCC}
\REQUIRE $T$ (number of epochs), $d$ (number of dimension of encoded discrete representation), $K$ (number of codewords in codebook), $L$ (number of encoding representation samples per data sample), batch size $N$, the temperature $\tau$, channel variance $\delta^2$, modulation function $h_{\text{m}}$ and the demodulation function $g_{\text{m}}$.
\WHILE{epoch $t=1$ to $T$}
    \STATE Sample a mini-batch of data samples $\{(\x^{(i)}, \y^{(i)})\}_{i=1}^{N}$.
    \WHILE{$i=1$ to $N$}
        \WHILE{$j=1$ to $d$}
            \STATE Compute the logits $\boldsymbol{\ell}_j$ with (\ref{eq:logits}).
            \STATE Compute conditional distribution $p_{\bPhi}(z_j|\x^{(i)})$ 
            with (\ref{eq:posterior}).
            \STATE Compute the conditional entropy term $H_{\bPhi}(Z_j|\x^{(i)})$.
            \STATE Sample the discrete representation $\{z_j^{(i,l)}\}_{l=1}^L$ using the Gumbel-softmax trick in (\ref{eq:gumbel_max}) and compute the entropy terms $\{H(\hat{Z}_j|z_j^{(i,l)})\}_{l=1}^L$.
        \ENDWHILE
        \STATE Sample the noise $\{\boldsymbol{\epsilon}^{(i,l)}\}_{l=1}^L \sim \mathcal{CN}(0, \sigma^2\mathbf{I})$.
        \STATE Estimate the corrupted discrete representation $\{\hz^{(i,l)}\}_{l=1}^L$.
    \ENDWHILE
    \STATE Compute the loss $\tilde{\mathcal{L}}_{\textrm{VRIB}}(\bPhi, \bTheta)$ based on (\ref{eq:RIB_MC}).
    \STATE Update the parameters $\bPhi$ and $\bTheta$ (including $\mathbf{M}$) through backpropagation.
\ENDWHILE
\end{algorithmic}
\end{algorithm}
        \subsubsection{Digital Modulation}
	To transmit the encoded discrete representation $\z$ under the channel models, the digital modulation scheme is used to map each discrete dimension $z_j$ into the constellation symbols. We propose to couple the discrete space with the constellation of modulation. Specifically, note that each discrete dimension $z_j\in[K]$ is $K$-way categorical, we adopt the digital modulation with $K$-points constellation design. Formally, each discrete dimension $z_j$ is mapped into a complex-valued channel symbol by a modulation function $h_{\text{m}}:[K] \to \mathbb{C}$. 
 The following AWGN channel model is simulated by a non-trainable physical layer with the injection of the Gaussian noise $\boldsymbol{\epsilon} \sim \mathcal{CN}(\mathbf{0}, \sigma^2\mathbf{I})$.
 On the receiver side, the demodulator demodulates the noisy complex-valued symbol into the corrupted discrete representation $\hat{z}_j$ using a demodulation function $g_{\text{m}}: \mathbb{C} \to [K]$. Each corrupted discrete dimension $\hat{z}_j$ can be formally obtained by:
    	\begin{align}
		\hat{z}_j &= g_{\text{m}}(h_{\text{m}}(z_j)+ \epsilon_j). \label{eq:extended_channel}
	\end{align}
 Note that the conditional distribution $p(\hz|\z)$ is determined by $h_{\textrm{m}}$, $g_{\textrm{m}}$ and noise variance $\sigma^2$, the entropy $H(\hat{Z}_j|z_j)$ can be computed for different discrete value of $z_j$, and the entropy term $H(\hat{Z}|\z)$ is the summation of $H(\hat{Z}_j|z_j)$ as presented in (\ref{eq:entropy_z}).
 In the training stage, the maximization of $\bE_{p_{\bPhi}(\z|\x)}[H(\hZ|\z)]$ in $\mathcal{L}_{\text{VRIB}}(\bPhi, \bTheta)$ optimizes the encoder $p_{\bPhi}(z_j|\x)$ to minimize the transmission rate and induce the optimal marginal distribution $p_{\bPhi}(z_j)$ for the digital modulation in the DT-JSCC systems. 
 
 The training procedures for DT-JSCC are summarized in Algorithm~\ref{algo-DTJSCC}.

	\subsection{Practical Advantages of DT-JSCC}
        With the RIB framework, our proposed task-oriented communication scheme DT-JSCC owns several practical advantages, highlighted as follows. 
	\begin{enumerate}
            \item \textbf{Robustness to channel fluctuation.} As outlined in Section~\ref{sec:principle}, the RIB framework is developed to control the coded redundancy for robust encoding. When applied to DT-JSCC systems, the level of coded redundancy depends on the degree of uncertainty of the encoded representation $\z$. Specially, the small value of $\beta$ in $\mathcal{L}_{\text{RIB}}(\bPhi, \bTheta)$ encourages the minimization of $H_{\bPhi}(\hZ_j|\x)$, leading to the one-hot categorical $p_{\bPhi}(\hat{Z_j}|\x)$, where only one codeword can be utilized for correct inference result. Whereas the large $\beta$ induces the distribution $p_{\bPhi}(\hat{z}_j|\x)$ with large entropy and multiple codewords are acceptable for the following inference model $q_{\bTheta}(\y|\hz)$ to output correct inference results.
            \item \textbf{Compatibility with modern mobile systems.} The proposed DT-JSCC method is well-suited for modern mobile systems. Modern mobile networks such as 5G are worldwide deployed with digital modulation instead of analog modulation. 
             Without extra efforts to transmit the continuous feature vectors, our proposed DT-JSCC method can be deployed in modern wireless networks based on affordable digital modulation schemes. Specifically, the high-resolution modulation for continuous value transmission brings a huge burden to the hardware of communication systems. But, in DT-JSCC communication systems, the modulation with finite-points constellation design makes the discrete representation transmission affordable for the capacity-limited communication systems.
		\item \textbf{Extendability to other AI applications. } 
  The information capacity of discrete representations is intrinsically limited by the cardinality of discrete alphabets. The compactness of discrete representation is beneficial in modeling perceptual signals. For instance, the semantic information of images or audio can be effectively represented by a sequence of discrete symbols. This advantage presents opportunities for the future adaptation of DT-JSCC in generative modeling tasks such as speech synthesis\cite{baevski2020wav2vec}, image generation\cite{van2017neural}, and text-to-image generation\cite{razavi2019generating}. Such capabilities would be particularly advantageous for serving AI applications including digital twin\cite{tao2018digital} and metaverse\cite{maier2020internet} in future 6G. 

	\end{enumerate}
\section{Experiments}
	\label{sec:exp}
	In this section, we present the experimental evaluations of the inference performance and robustness of the proposed DT-JSCC methods on image classification benchmarks compared with the state-of-the-art task-oriented communication schemes. Then, the ablation studies are conducted to verify that the RIB framework can achieve an optimal informativeness-robustness tradeoff, and investigate the effect of codebook size on the inference performance of DT-JSCC\footnote{The source code is available at https://github.com/SongjieXie/Discrete-TaskOriented-JSCC. }.
	\subsection{Datasets}
	The experiments are conducted on two benchmark datasets with random realizations of the channel under consideration, including \emph{MNIST} \cite{lecun1998gradient} and \emph{CIFAR-$10$} \cite{krizhevsky2009learning}.
	
	The MNIST dataset is a set of handwritten character digits from $0$ to $9$. It contains a training set of $60,000$ gray-scale images and a test set of $10,000$ sample images. 
	
	The CIFAR-10 dataset is a set of $32\times32$ color images in $10$ classes. It contains $50,000$ training images with $5,000$ images per class and $10,000$ test images. In the following experiments, we apply data augmentation to the CIFAR-10 dataset in the training process of neural networks. The augmentation policy consists of random cropping and horizontal flipping.
	\subsection{Experimental Settings}\label{subsec:exp_setting}
	\subsubsection{Compared Methods}
	We select two state-of-the-art task-oriented communication schemes as baseline methods for performance comparison, including \emph{DeepJSCC} \cite{bourtsoulatze2019deep, jankowski2020wireless} and \emph{Varitional Feature Encoding (VFE)} \cite{shao2021learning}. All the baseline methods are learning-based techniques with end-to-end JSCC architectures.
	\begin{itemize}
		\item DeepJSCC is initially proposed for data-oriented communication systems, which map the data to the channel input symbols by DNN-based encoders. 
		\item VFE is a learning-based scheme for task-oriented communication, which is optimized to reduce communication overhead using the variational Information Bottleneck principle. The task-relevant features are extracted by a variational encoder and are mapped into the symbols transmitted across noisy channels. By introducing sparsity-inducing prior distributions (i.e., the log-uniform distribution \cite{kingma2015variational}) and pruning on the zero dimensions, the symbols are transmitted over fewer dimensions to achieve low-latency communication for VFE.  
	\end{itemize}
	\subsubsection{Evaluations}\label{subsubsec:evaluations}
Our experiments on task-oriented communication center around classification tasks using the MNIST and CIFAR-10 benchmark datasets. As these datasets have balanced classes and a substantial number of classes, the standard metric top-$1$ error rate or top-$1$ accuracy are used to evaluate the classification results. A lower error rate or a higher accuracy value indicates superior performance in classification.
 
    We simply adopt the full-resolution constellation to modulate the continuous representation encoded by DeepJSCC and VFE, while for the proposed discrete method DT-JSCC, we use $K$-points constellation modulation $K$-PSK to modulate the discrete representations encoded by DT-JSCC.
    Note that the communication latency is determined by the dimension $d$ of the encoded representation. For a fair comparison, the representations encoded by all the evaluated methods have the same dimensionality $d=16$. Following the same setting used in \cite{shao2021learning} to set up a fixed bandwidth $12.5$kHz with symbol rate $9,600$ Baud, all the methods are evaluated with relatively low latency $t=1.67\text{ms}$ throughout the experiments.
	
	\subsubsection{Implementions}
	\begin{table}
		\caption{The Neural Network Architecture for MNIST Classification Task}
		\label{tab:arc-mnist}
		\begin{tabular}{p{0.2\columnwidth}|p{0.45\columnwidth}|p{0.2\columnwidth}}
			\toprule
			& \textbf{Layer} & \textbf{Outputs} \\
			\midrule
			\textbf{Encoder} & Dense $+$ Tanh / GVQ & $d$\\
			\midrule
			\multirow{3}{*}{\textbf{Inference}} & \text{Dense} $+$ \text{ReLU} & 1024\\
			& Dense $+$ ReLU & 256\\
			& Dense $+$ Softmax & 10\\
			\bottomrule
		\end{tabular}
	\end{table}
	\begin{table}
		\caption{The Neural Network Architecture for CIFAR-10 Classification Task}
		\label{tab:arc-cifar}
		\begin{tabular}{p{0.17\columnwidth}|p{0.45\columnwidth}|p{0.22\columnwidth}}
			\toprule
			& \textbf{Layer} & \textbf{Outputs} \\
			\midrule
			\multirow{3}{*}{\textbf{Encoder}} & $\textrm{Conv}\times 2 + \textrm{ResNet Block}$ & $128\times 16\times 16$\\
			& $\textrm{Conv}\times 2 + \textrm{ResNet Block}$ & $512\times 4 \times 4$\\
			& $[\textrm{Dense}+\textrm{Tanh}]$ / GVQ     & $d$\\
			\midrule
			\multirow{3}{*}{\textbf{Inference}}& $\textrm{ResNet Block}\times 2$  & $512\times 4 \times 4$\\
			& MaxPooling                       & $512$\\
			& Dense $+$ Softmax & 10\\
			\bottomrule
		\end{tabular}
	\end{table}
To ensure fair comparisons, we use the same neural network backbone with limited computation and memory resources on all local devices for both the proposed schemes and the compared methods. Since we are considering multi-class classification tasks, the inference network employs a neural network-based multi-class classifier, which includes a dense layer and a softmax layer.
	
 For the classification tasks for datasets MNIST and CIFAR-10, the corresponding backbones are designed as follows.
	\begin{itemize}
		\item For the experiments of the MNIST dataset, we adopt dense neural networks as the backbone shown in Table~\ref{tab:arc-mnist}. On the transmitter, the encoder is based on a single-layer neural network, and the multi-layer dense network is adopted as the inference model on the receiver. Besides, we use the tanh activate function to map the encoded features into $d$-dimensional continuous representations for DeepJSCC and VFE. For DT-JSCC, we adopt Gumbel-based vector quantization (GVQ) to map the encoded features to $d$-dimensional discrete representations. 
		\item For the experiments of the CIFAR-10 dataset, we use the convolutional layers and ResNet\cite{he2016deep} blocks to form the learning backbones for local encoders and the central inference model. The DT-JSCC quantizes the encoded features on local devices by Gumbel-based vector quantization (GVQ). The compared methods (i.e., DeepJSCC and VFE) map the encoded features into channel input symbols by adding a dense layer at the end of the backbone of the encoder. At the central server, the DT-JSCC leverages the transmitted discrete representation with the learned codebook, while the compared methods map the received symbols into feature tensors with an additional dense layer. The overall architecture of the neural network backbone is shown in Table~\ref{tab:arc-cifar}.
	\end{itemize}
 
 The AWGN channels are simulated by adding Guassian noise $\boldsymbol{\epsilon}\sim \mathcal{CN}(0, \sigma^2 \mathbf{I})$ to channel input signals with the channel PSNR defined as: 
	\begin{equation}
		\textrm{PSNR} = 10\log_{10}\frac{P_{\textrm{max}}}{\sigma^2}(\textrm{dB})
	\end{equation}
        where $P_{\textrm{max}}$ denote the maximal power of the channel input signal. 
	
	\subsection{Experimental Results}\label{subsec:exp_results}
	In the experiments, we evaluate the performance of the proposed DT-JSCC compared to the baselines with relatively low transmission latency. We set up the dimension of the encoded representation equal to $16$ for all the discrete and continuous methods.
    In the proposed DT-JSCC method, we use an acceptable codebook size $K=16$ and adopt a simple digital modulation scheme $16$-PSK in the end-to-end system. Furthermore, the hyperparameter $\beta \in [10^{-4}, 10^{-1}]$ is determined by using the grid search to find the optimal $\beta$ for the CIFAR-10 and MNIST classification tasks.\footnote{For the MNIST and CIFAR-10 classification tasks, the target variable $Y\in \{0,1,\ldots,9\}$ limits the task-relevant information by its entropy $I(Y;\hat{Z}) \leq H(Y) \leq \log 10$, which indicates a small range compared to the amount of coded redundancy $R(\bPhi) \leq H(\hat{Z}) \leq 64$ that may be present in $\hz$. Therefore, a small $\beta$ can effectively control the informativeness-robustness tradeoff, as formulated by the RIB framework.}
    For VFE, the initial dimension is set by $20$ and we adjust the sparsity-control parameter to reduce the dimension from $20$ to $16$. 
	
	\subsubsection{Inference Performance}
	\begin{table}
		\centering
		\caption{The inference accuracy of evaluated methods for the MNIST classification task.  }
		\label{tab:res-mnist}
		\resizebox{.97\linewidth}{!}{
			\begin{tabular}{c|ccccc}
				\toprule
				PSNR & 4 dB & 8 dB & 12 dB & 16 dB & 20 dB\\
				\midrule
				DeepJSCC & 86.63 &93.92& 95.39& 95.63& 95.91\\
				VFE   &    86.69& 93.95& 95.41& 95.79& 96.03 \\
				DT-JSCC&  \textbf{96.66}&\textbf{97.21}& \textbf{97.25}& \textbf{97.72} & \textbf{97.93}\\
				\bottomrule
		\end{tabular}}
	\end{table}
	\begin{table}
		\caption{The inference accuracy of evaluated methods for the CIFAR-10 classification task.}
		\label{tab:res-cifar}
		\resizebox{0.97\linewidth}{!}{
			\begin{tabular}{c|ccccc}
				\toprule
				PSNR & 4 dB & 8 dB & 12 dB & 16 dB & 20 dB\\
				\midrule
				DeepJSCC &91.22& 91.66& 91.80 & 91.90 &91.93 \\
				VFE&      91.33& 91.67& 91.84 & 91.94 &91.98 \\
				DT-JSCC&  \textbf{91.46}&\textbf{91.93}& \textbf{91.91}& \textbf{92.26} & \textbf{92.14}\\
				\bottomrule
		\end{tabular}}
	\end{table}
 \begin{figure*}
		\centering
		\subfloat[$\textrm{PSNR}_{\textrm{train}}= 8$ dB]{
			\centering
			\includegraphics[width=0.3\linewidth]{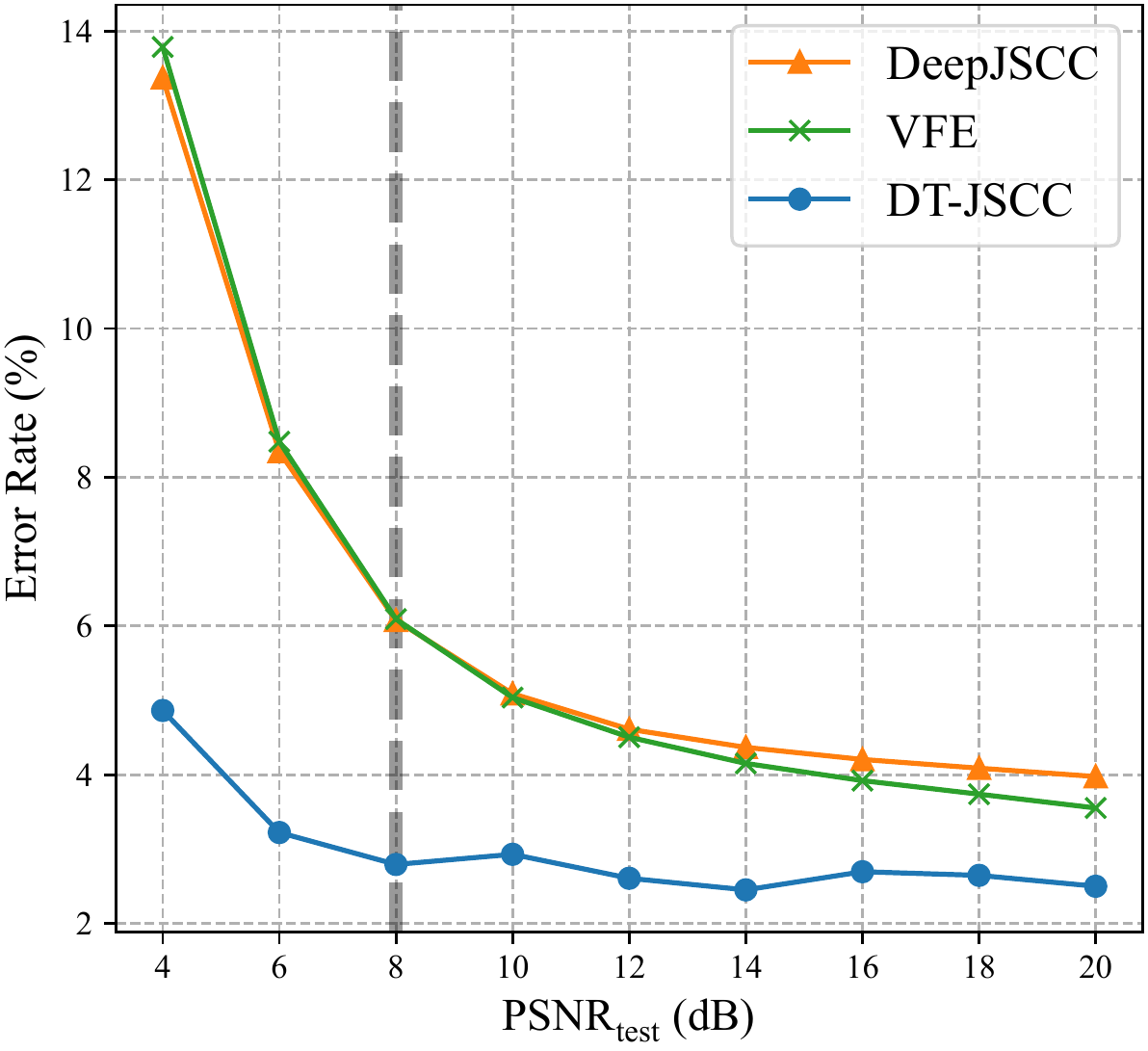}
		}
		\subfloat[$\textrm{PSNR}_{\textrm{train}}= 12$ dB]{
			\centering
			\includegraphics[width=0.3\linewidth]{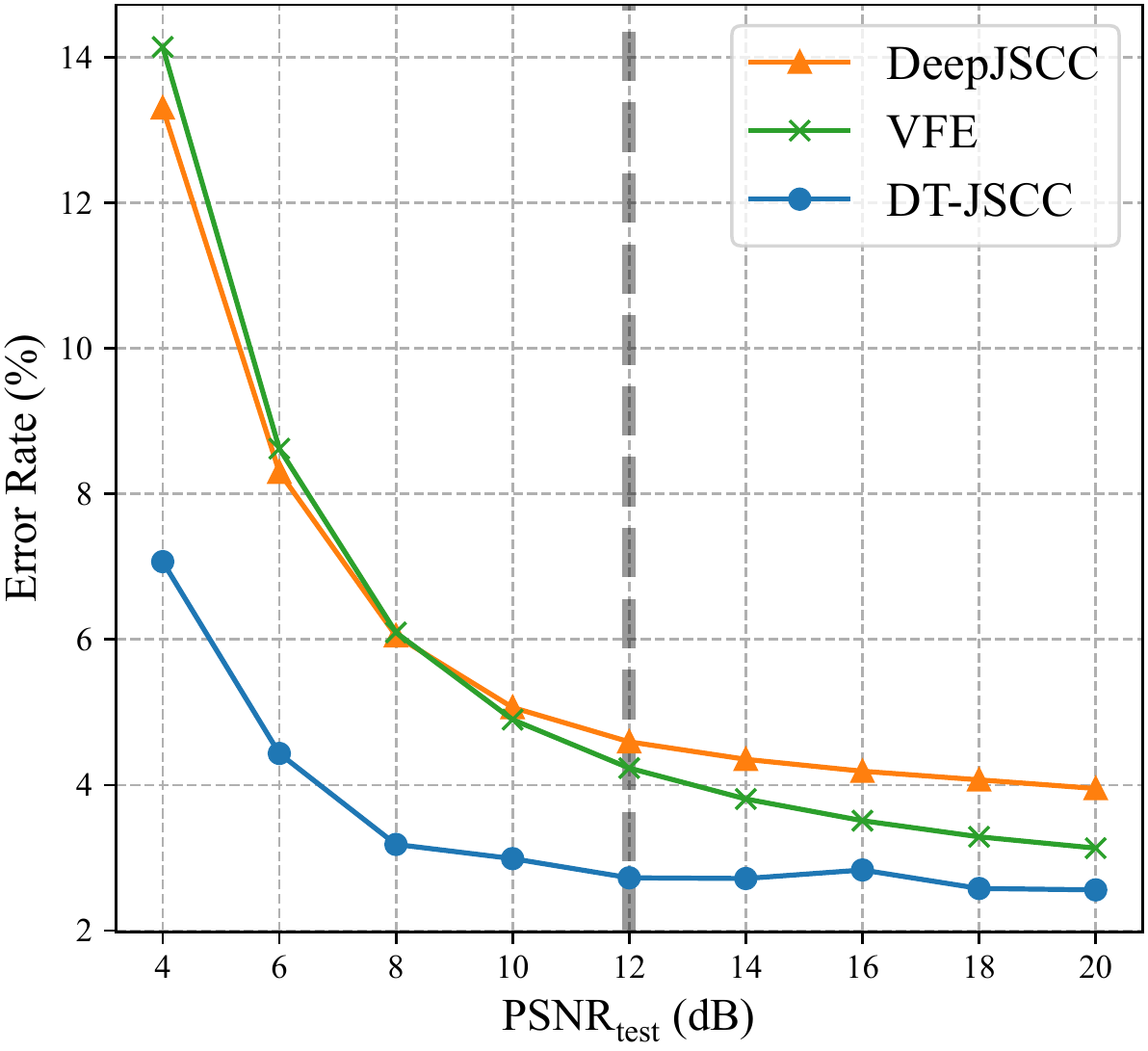}
		}
		\subfloat[$\textrm{PSNR}_{\textrm{train}}= 16$ dB]{
			\centering
			\includegraphics[width=0.3\linewidth]{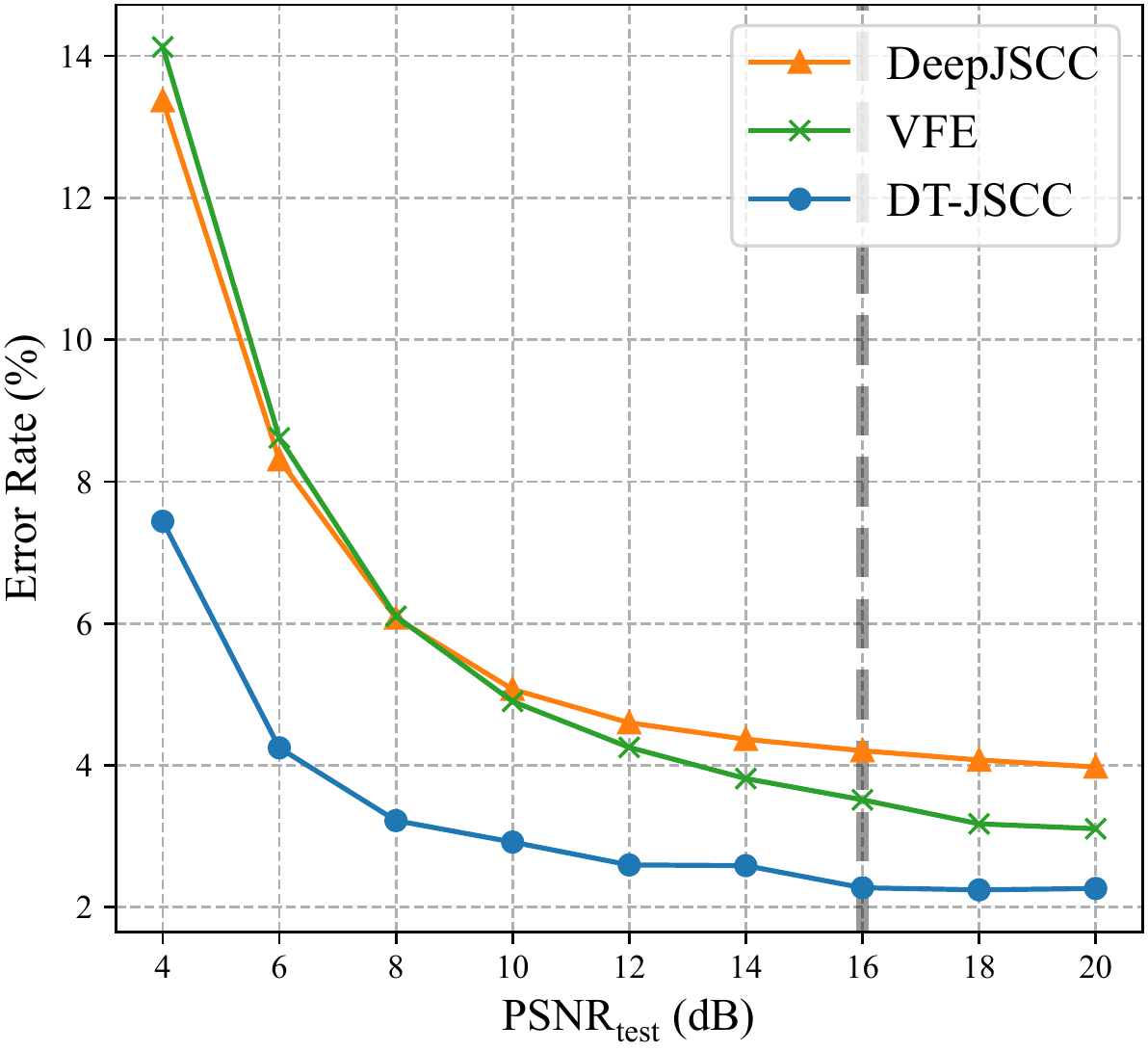}
		}
		\caption{Performance of DT-JSCC and the compared methods on MNIST classification task over testing AWGN channels with $\textrm{PSNR}_{\textrm{test}} \in [4\textrm{dB}, 20\textrm{dB}]$. Each method is trained over AWGN channels with a specific PSNR value, (a) $\textrm{PSNR}_{\textrm{train}}= 8\textrm{dB}$, (b) $\textrm{PSNR}_{\textrm{train}}= 12\textrm{dB}$ and (c) $\textrm{PSNR}_{\textrm{train}}= 16\textrm{dB}$, which is marked by a vertical dash line.
		}
		\label{fig:exp-mnist-com}
	\end{figure*}
	\begin{figure*}
		\centering
		
		\subfloat[$\textrm{PSNR}_{\textrm{train}}= 8$ dB]{
			\centering
			\includegraphics[width=0.3\linewidth]{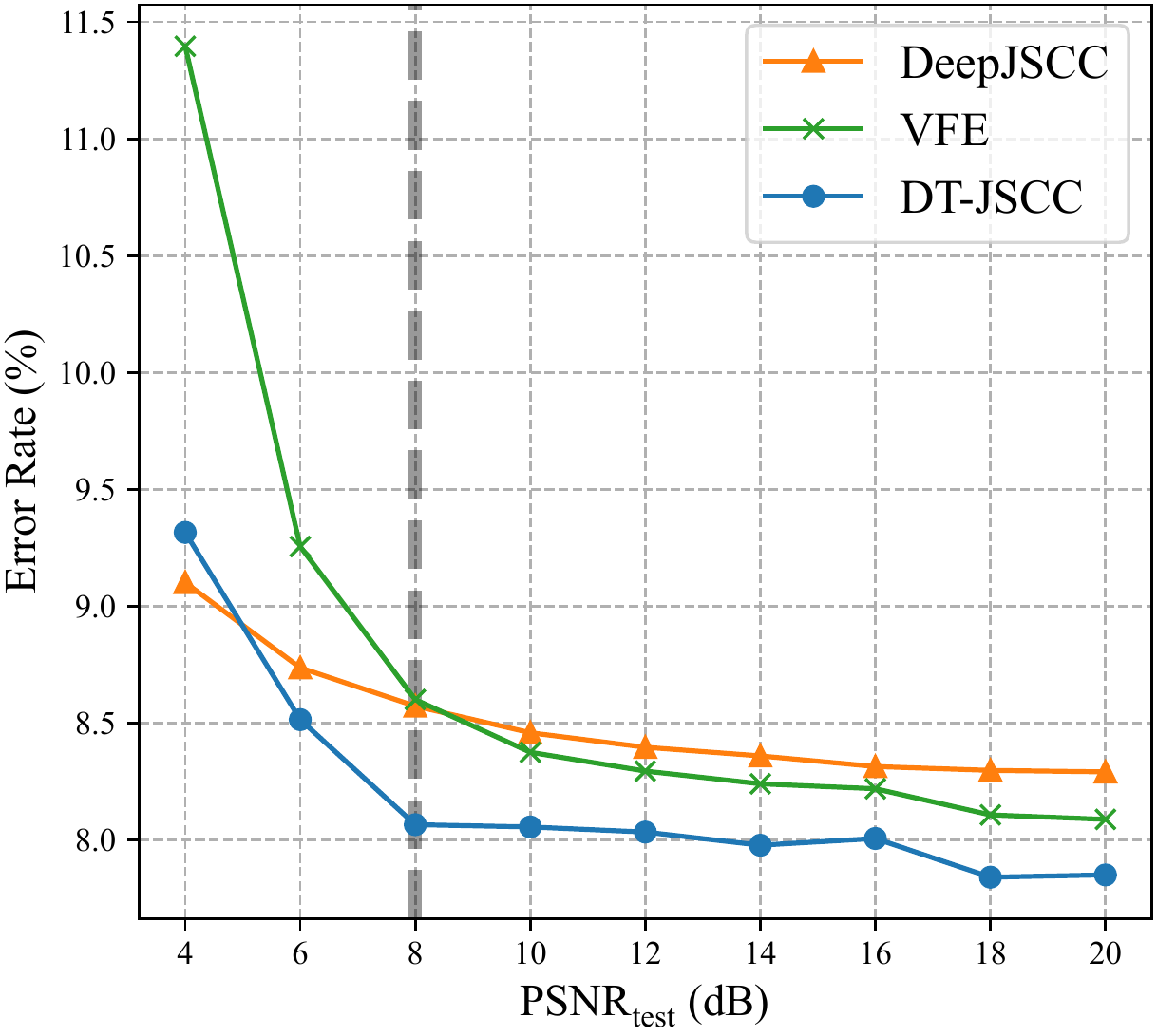}
		}
		\subfloat[$\textrm{PSNR}_{\textrm{train}}= 12$ dB]{
			\centering
			\includegraphics[width=0.3\linewidth]{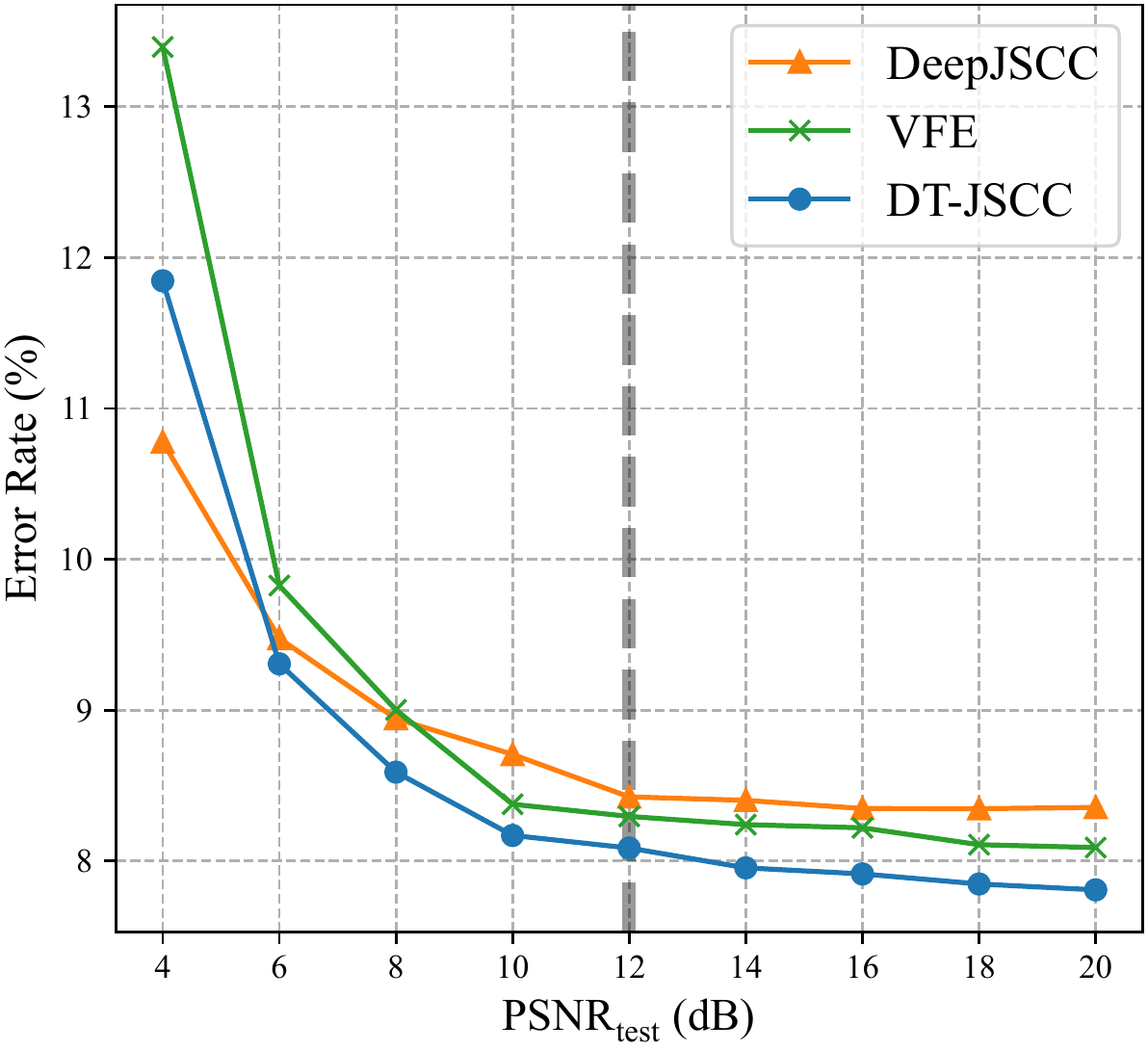}
		}
		\subfloat[$\textrm{PSNR}_{\textrm{train}}= 16$ dB]{
			\centering
			\includegraphics[width=0.3\linewidth]{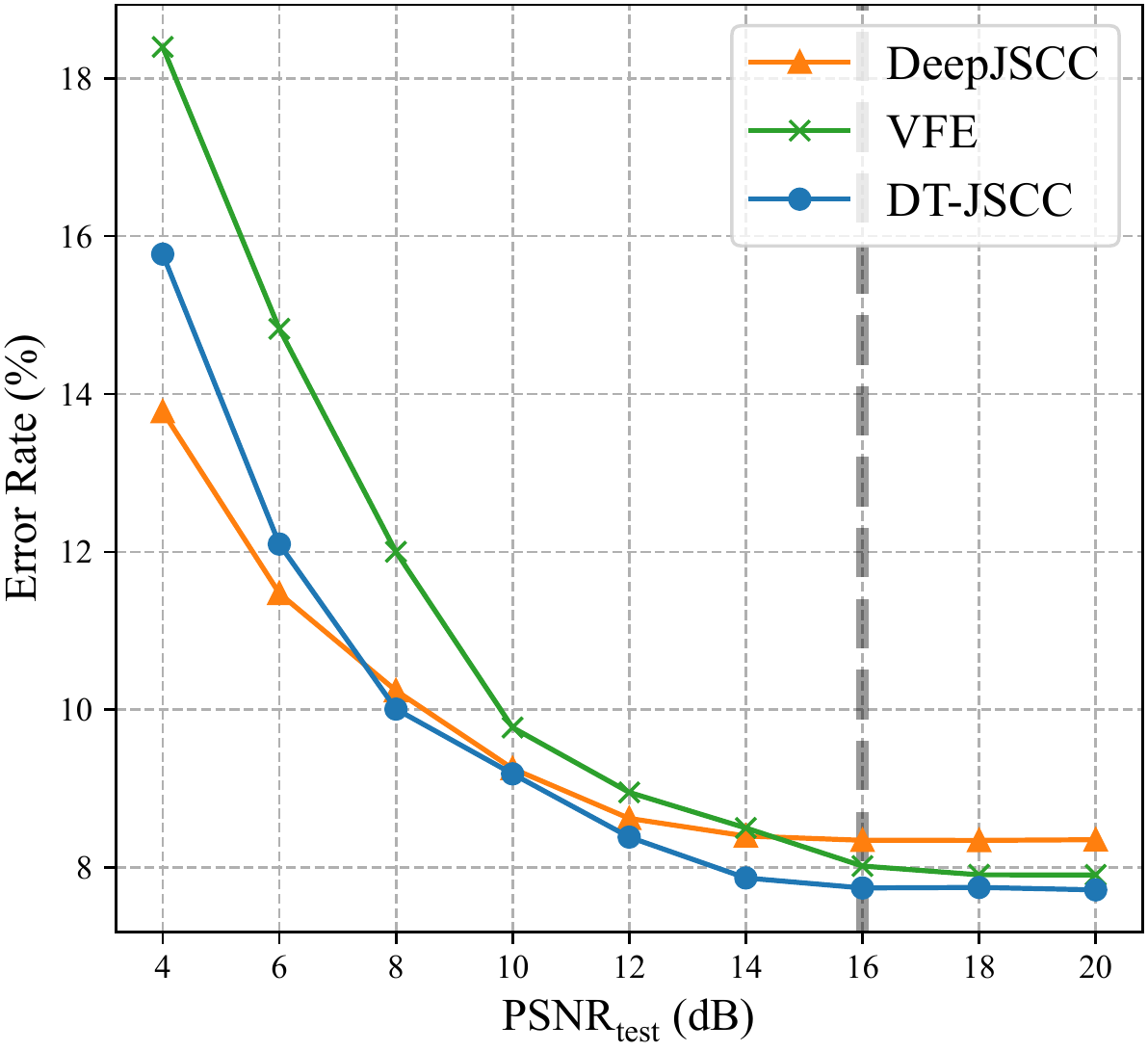}
		}
		
		\caption{Performance of DT-JSCC and the compared methods on CIFAR-10 classification task over testing AWGN channels with $\textrm{PSNR}_{\textrm{test}} \in [4\textrm{dB}, 20\textrm{dB}]$. Each method is trained over AWGN channels with a specific PSNR value, (a) $\textrm{PSNR}_{\textrm{train}}= 8\textrm{dB}$, (b) $\textrm{PSNR}_{\textrm{train}}= 12\textrm{dB}$ and (c) $\textrm{PSNR}_{\textrm{train}}= 16\textrm{dB}$, which is marked by a vertical dash line.
		}
		\label{fig:exp-cifar-com}
	\end{figure*}
	We first evaluate the cooperative inference performance of the proposed DT-JSCC with the ideal assumption that the channel models at the training and testing stage have the same PSNR. 
 Table~\ref{tab:res-mnist} presents the inference accuracy of the evaluated methods on the MNIST dataset for classification tasks. Our proposed method, DT-JSCC, outperforms the two baseline methods in all cases, demonstrating the superiority of the discrete representation encoded by DT-JSCC. The significant increase in accuracy confirms that our method successfully encodes informative messages that are shared between the transmitter and the receiver.
Interestingly, both DeepJSCC and VFE achieve similar accuracies, suggesting that  the continuous encoded representation may reach its limit in inference performance when the dimension is low.

 Moving on to the CIFAR-10 dataset, Table~\ref{tab:res-cifar} shows that our method performs better than the baselines, especially when the PSNR is less than 8 dB. It is worth noting that VFE outperforms DeepJSCC because it identifies and reduces the redundant dimensions in the encoded continuous representation. However, the discrete representation of DT-JSCC contains more informative messages due to the learned codebook, leading to superior performance.
In summary, our proposed method, DT-JSCC, achieves better inference performance in both datasets, demonstrating the effectiveness of discrete representation encoding in task-oriented communication. 
	\subsubsection{Robustness}
	Next, we validate the robustness of the proposed DT-JSCC scheme to the variations in channel quality. All the evaluated methods are end-to-end trained for a specific channel PSNR, denoted as $\text{PSNR}_{\text{train}} \in \{8\textrm{dB}, 12\textrm{dB}, 16\textrm{dB},\} $, and then deployed in the AWGN channel model with the testing PSNR changing from $4$dB to $20$dB, denoted as $\text{PSNR}_{\text{test}} \in [4\textrm{dB}, 20\textrm{dB}]$.

 The experimental results shown in Fig.\ref{fig:exp-mnist-com} and Fig.\ref{fig:exp-cifar-com} indicate the inference performance of the evaluated methods for different channel quality conditions from those used during the training stage. We observe that the inference accuracy improves gradually and then saturates with the increase of $\textrm{PSNR}_{\textrm{test}}$. Interestingly, we found that the DeepJSCC outperforms the VFE when the channel conditions are worse than those used for training, i.e., $\textrm{PSNR}_{\text{test}}< \textrm{PSNR}{\text{train}}$. Conversely, when $\textrm{PSNR}_{\text{test}} > \textrm{PSNR}_{\text{train}}$, the VFE performs better than the DeepJSCC, confirming our insight that reducing redundancy in the encoded representation degrades robustness to channel quality fluctuations.
  
 For the MNIST classification task, our proposed DT-JSCC method consistently outperforms the two baselines with a significant increase in accuracy, demonstrating the effectiveness of our proposed approach. In addition, we observe that our method exhibits better gradual performance degradation than VFE and DeepJSCC when the channel quality in the test phase is worse than that in the training phase, as indicated by $\text{PSNR}_{\text{test}} < \text{PSNR}_{\text{train}}$. This indicates that our proposed method is more robust than the baselines in handling channel quality variations. The improved robustness of our method can be attributed to the combination of discrete representation and coded redundancy introduced by the RIB framework, which enables our method to better tolerate the channel noise and fluctuations.

 The experimental results on the CIFAR-10 image classification task demonstrate the superior performance of our proposed method compared to the two baselines for the channel models with $\textrm{PSNR}{\textrm{test}}$ in the $[8 \textrm{dB}, 20\textrm{dB}]$ range. Interestingly, we found that the DeepJSCC achieves the best inference accuracy for the channel model with low testing PSNR, i.e., $\textrm{PSNR}{\textrm{test}} < 6 \textrm{dB}$, but the inference performance in the saturation region is much worse than that of the VFE and DT-JSCC. This tradeoff between informativeness and robustness is consistent with the principle presented in Section~\ref{sec:principle}. Our proposed methods exhibit more robustness than VFE to channel fluctuations and perform better than the two baselines when $\textrm{PSNR}_{\textrm{test}} > \textrm{PSNR}_{\textrm{train}}$. The proposed DT-JSCC methods leverage the compactness of discrete representation and coded redundancy introduced by the RIB framework to achieve high performance in good channel conditions and robustness to variations in channel quality, to a certain extent. Overall, these results demonstrate the effectiveness of our proposed DT-JSCC method for challenging image classification tasks.
 
	\subsection{Ablation Study}
	\label{subsec:ablation}
	\begin{figure}[t]
		\centering
		\includegraphics[width=7cm]{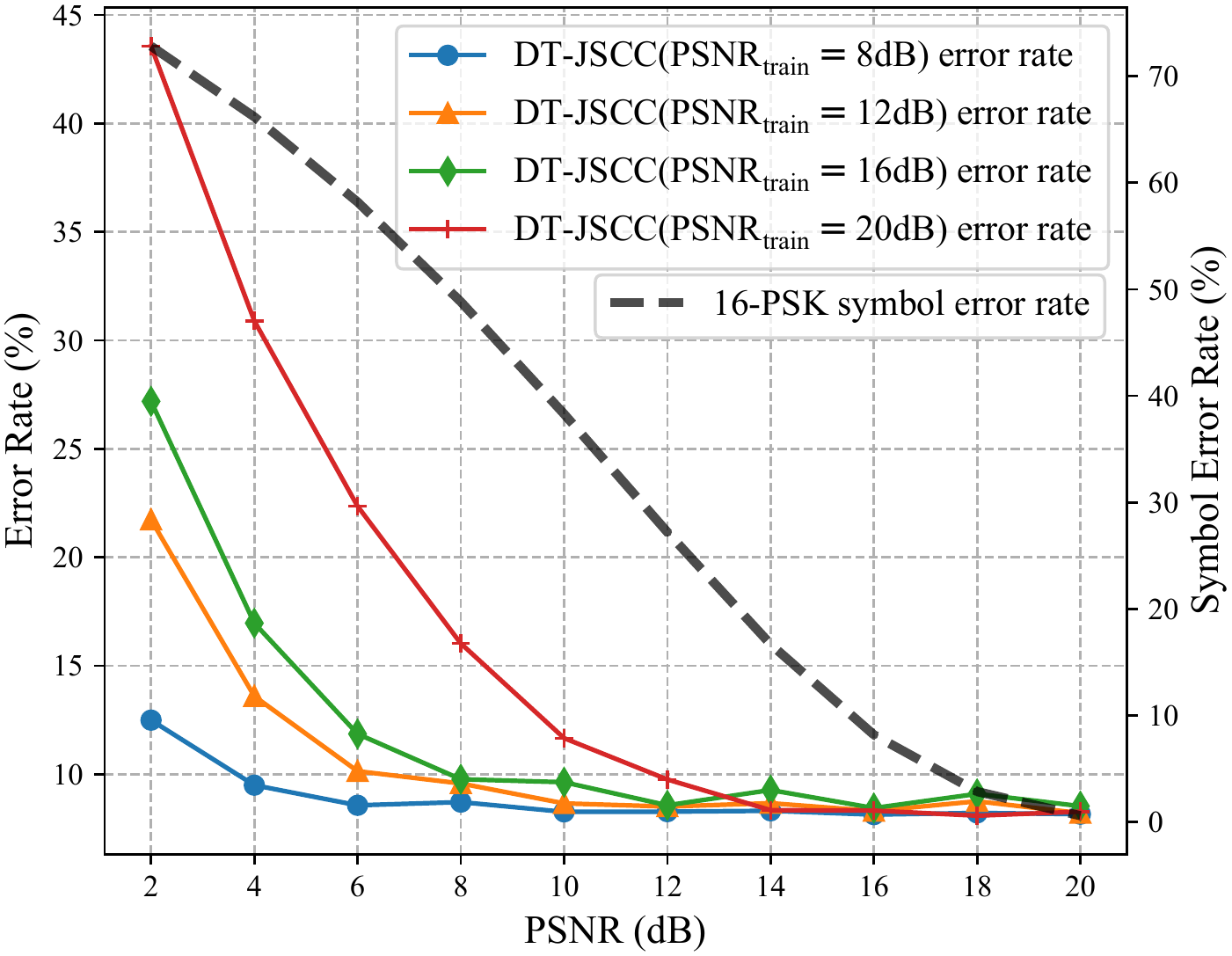}
		\caption{Performance comparison of DT-JSCC on CIFAR-10 classification task with respect to the training PSNR, i.e, $\textrm{PSNR}_{\textrm{train}} \in \{8\textrm{dB}, 12\textrm{dB}, 16\textrm{dB}, 20\textrm{dB}\}$. Each curve is obtained by training with the $\mathcal{L}_{\textrm{RIB}}(\bPhi, \bTheta)$ where $\beta = 0$. The gray dash curve is the symbol error rate of $16$-PSK modulation scheme.}
		\label{fig:ablation-channel}
	\end{figure}
	\begin{figure}
		\centering
		\subfloat[$\textrm{PSNR}_{\textrm{train}}= 8$ dB]{
			\centering
			\includegraphics[width=0.47\linewidth]{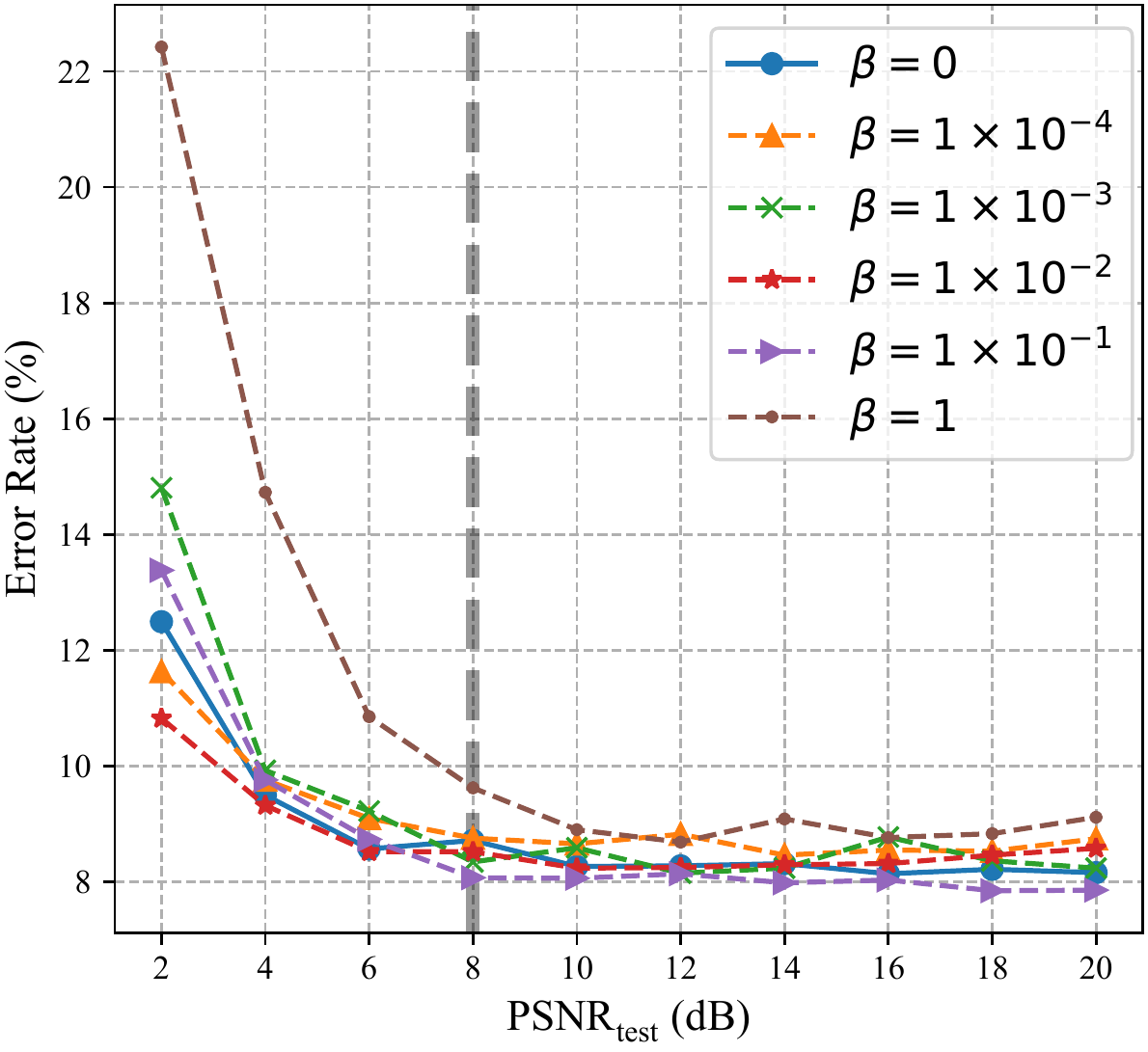}
		}
		\subfloat[$\textrm{PSNR}_{\textrm{train}}= 12$ dB]{
			\centering
			\includegraphics[width=0.47\linewidth]{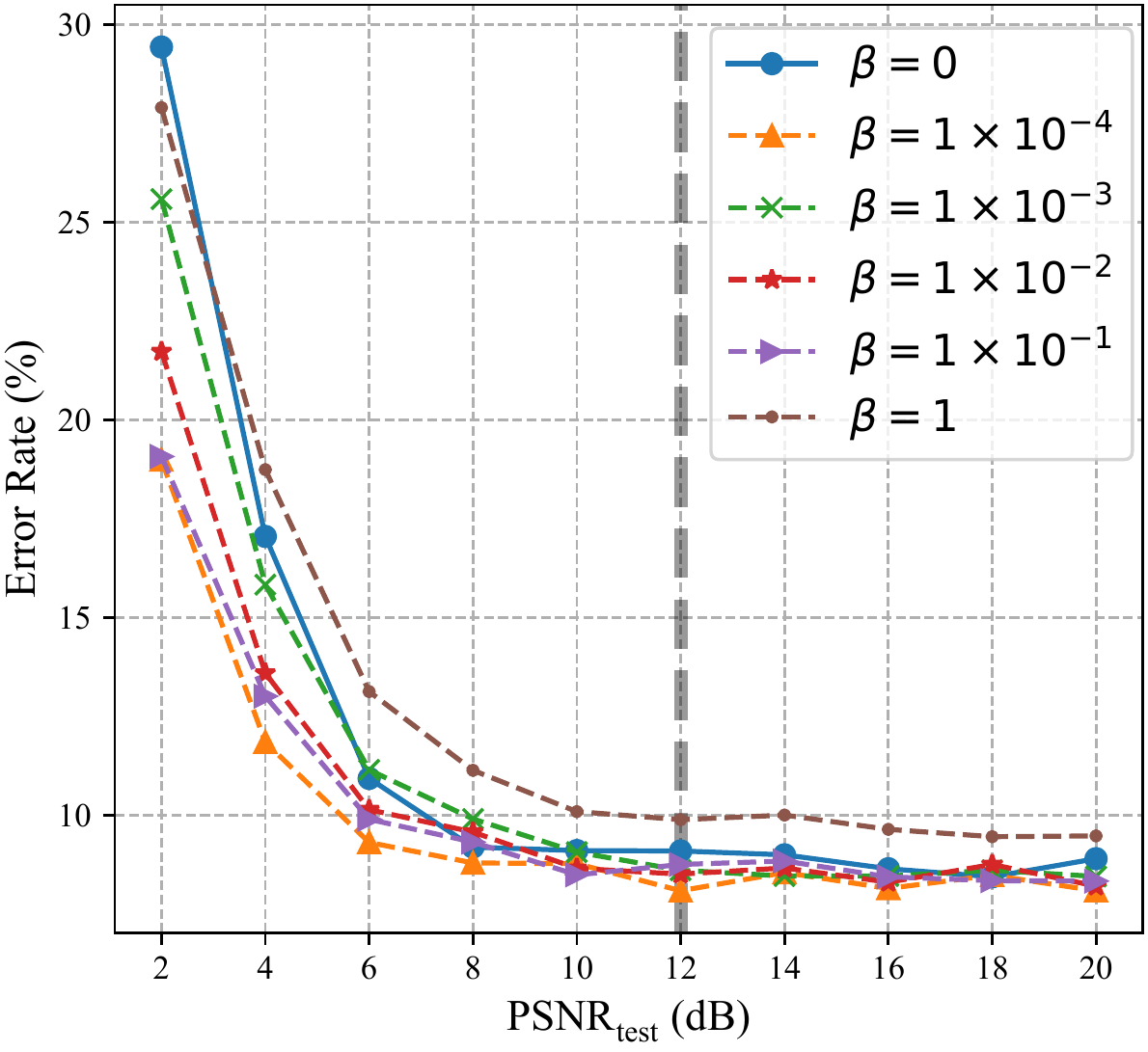}
		}
		\vfill
		\subfloat[$\textrm{PSNR}_{\textrm{train}}= 16$ dB]{
			\centering
			\includegraphics[width=0.47\linewidth]{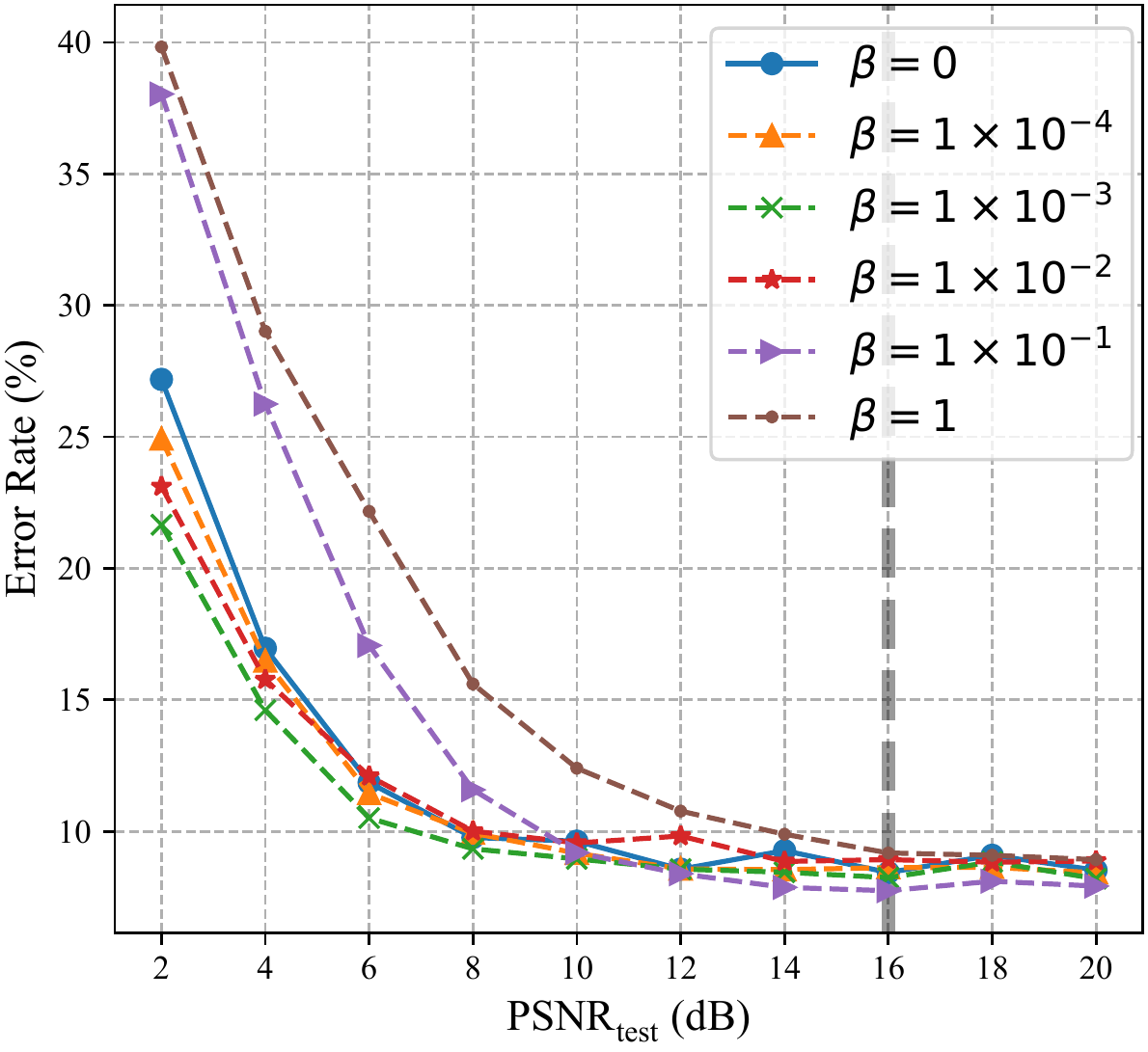}
		}
		\subfloat[$\textrm{PSNR}_{\textrm{train}}= 20$ dB]{
			\centering
			\includegraphics[width=0.47\linewidth]{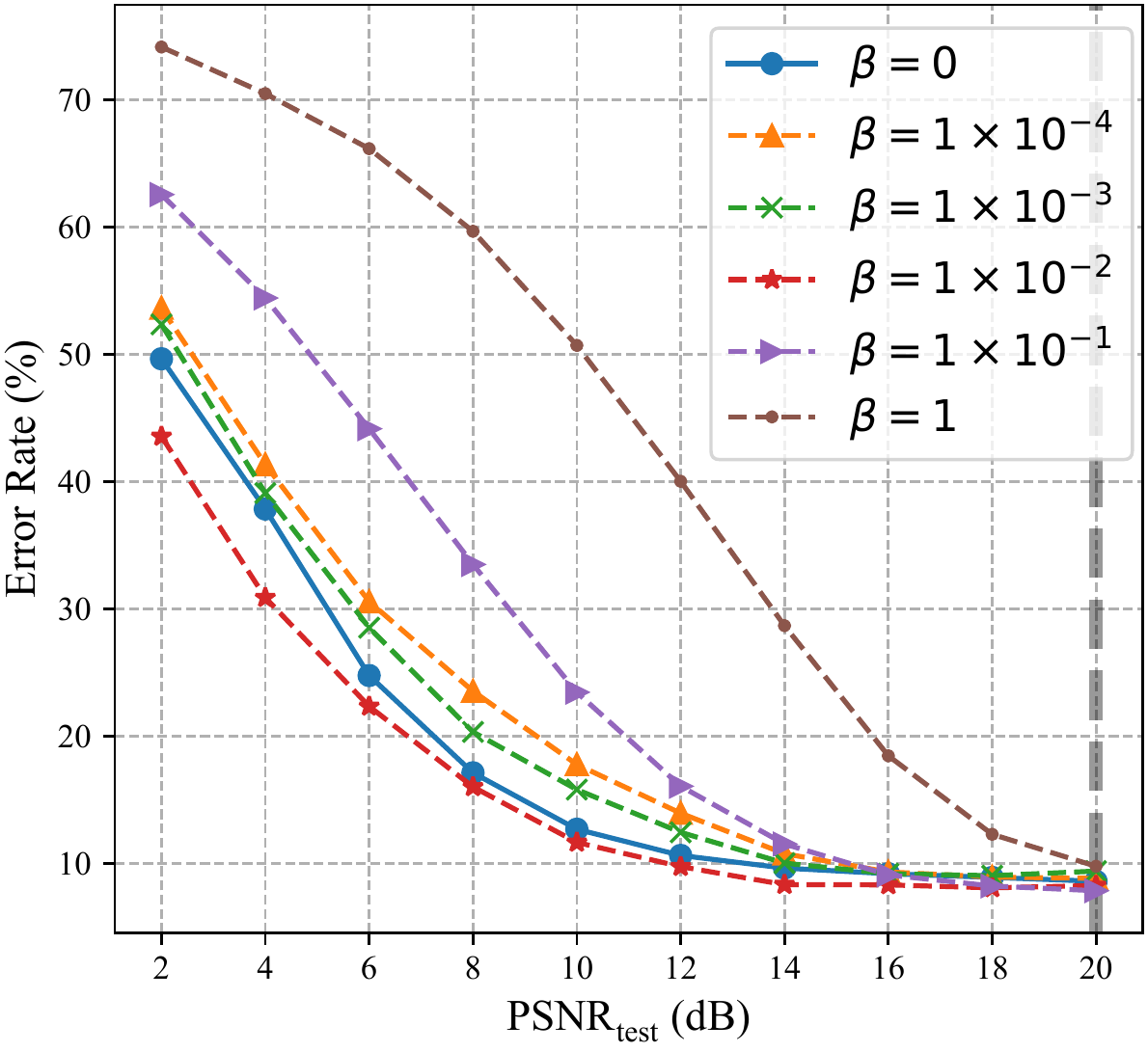}
		}
		\caption{Performance of the DT-JSCC with different $\beta \in [0, 1]$ trained under variant channel PSNR, (a) $\textrm{PSNR}_{\textrm{train}}= 8\textrm{dB}$, (b) $\textrm{PSNR}_{\textrm{train}}= 12\textrm{dB}$, (c) $\textrm{PSNR}_{\textrm{train}}= 16\textrm{dB}$ and (d) $\textrm{PSNR}_{\textrm{train}}= 20\textrm{dB}$.
		}
		\label{fig:ablation-lam}
	\end{figure}
	\begin{figure}
		\centering
		\subfloat[]{
			\centering
			\includegraphics[width=0.47\linewidth]{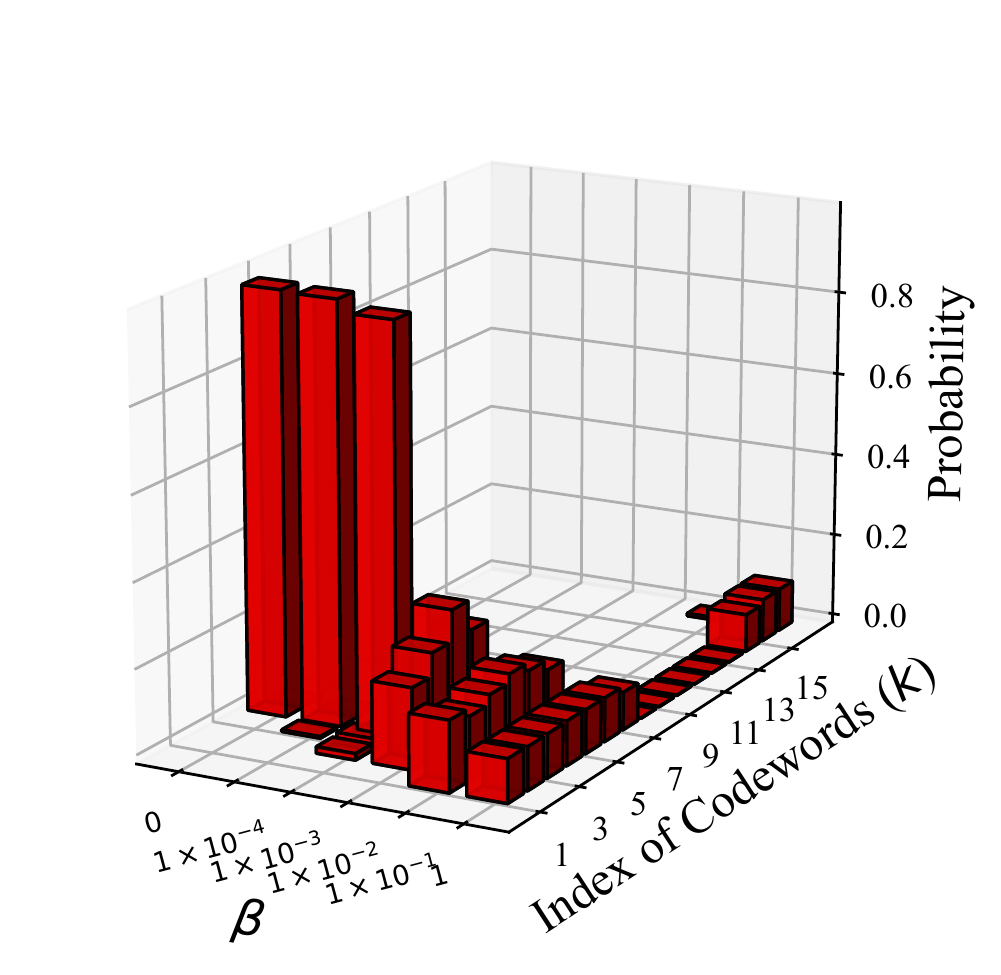}
		}
		\subfloat[]{
			\centering
			\includegraphics[width=0.47\linewidth]{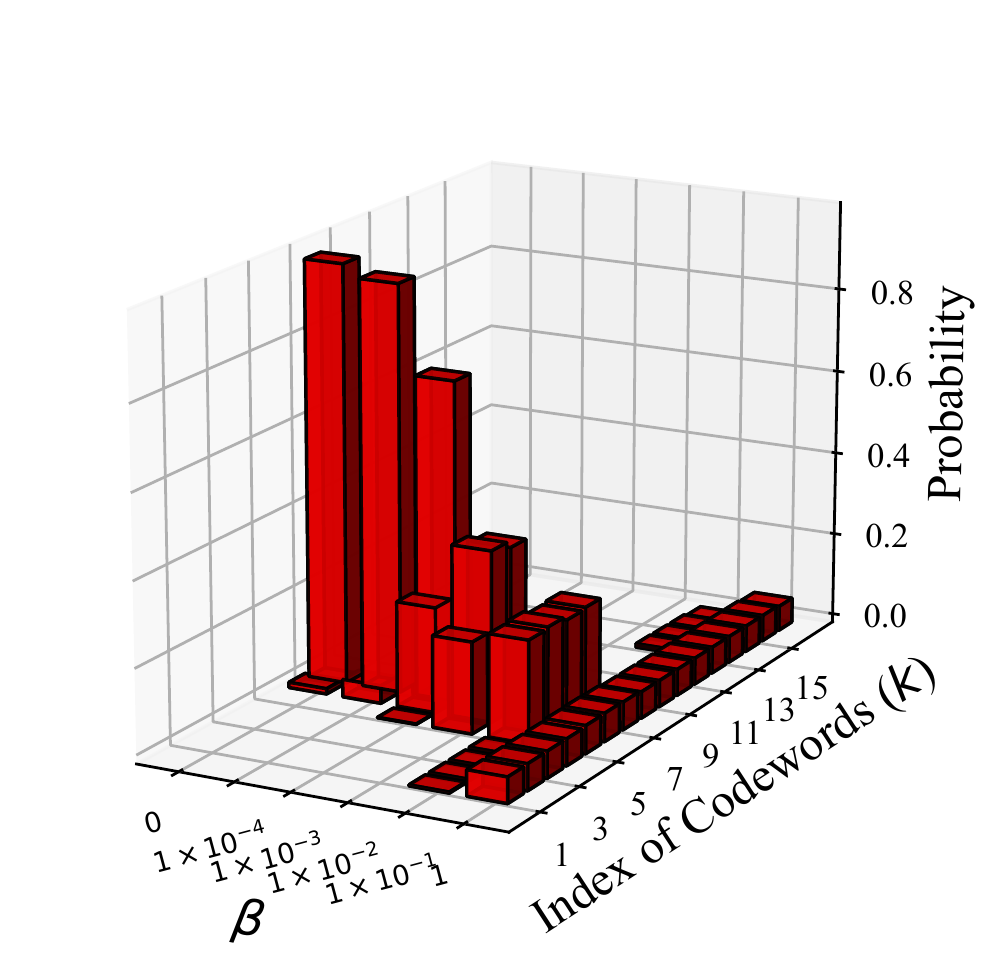}
		}
		\caption{Probability mass function of the categorical distribution $p_{\bPhi}(Z_0=k|\x)$ learned with different $\beta\in [0, 1]$ in the objective $\mathcal{L}_{\textrm{RIB}}(\bPhi, \bTheta)$ for a sample image in (a) MNIST and (b) CIFAR-10.
		}
		\label{fig:ablation-pmf}
	\end{figure}
	To evaluate the effectiveness of our proposed method, we conducted ablation studies on the robust encoding of the RIB framework by varying the redundancy parameter $\beta\in [0, 1]$ to control the amount of redundancy encoded in $\z$. Additionally, we conducted an ablation analysis to investigate the impact of the feature splitting strategy and the codebook size $K$ on the inference performance of the proposed DT-JSCC.
	\subsubsection{Robust Encoding in RIB Framework}\label{subsubsec:RobustEncoding}
	We firstly evaluate the DT-JSCC methods with $\beta=0$ in the objective $\mathcal{L}_{\text{RIB}}(\bPhi, \bTheta)$, where the coded redundancy is minimized to keep maximal task-relevant information. As depicted in Fig.~\ref{fig:ablation-channel}, each curve represents the inference error rate the DT-JSCC method optimized with specific AWGN channel PSNR. We can observe that the inference error rate of each method remains low even if symbol error rate of modulation is increasing, especially for the methods trained with low $\textrm{PSNR}_{\textrm{train}}$. The end-to-end training with the noise injection not only implicitly achieves the channel coding, but also exhibits certain robustness to variations in channel conditions.
	
	To justify the robust encoding of the RIB framework, we further compare the performance of proposed DT-JSCC methods with different $\beta \in [0,1]$. Fig.~\ref{fig:ablation-lam} shows that the DT-JSCC methods with certain $\beta >0 $ outperform the DT-JSCC with $\beta=0$ on the aspects of inference accuracy and the robustness to the channel fluctuation, implying that the coded redundancy controlled by $\beta$ can effectively improve the robustness. 
        We further investigate the coded redundancy encoded in the representation $\z$ by presenting the categorical distribution $p_{\bPhi}(Z_i=k|\x)$ of the DT-JSCC methods with different $\beta \in [0, 1]$. Fig.~\ref{fig:ablation-pmf} depicts the probability mass function $p_{\bPhi}(Z_0=k|\x)$ with different $\beta$ for MNIST and CIFAR-10 classification tasks. We can observe that the objective $\mathcal{L}_{\textrm{RIB}}(\bPhi, \bTheta)$ with larger $\beta$ induces the denser probability mass function, which allows more discrete distortion of $Z_i$ introduced by the channel noise and makes the encoded discrete representation $\z$ more robust to the channel noise. On the other hand, the excessive redundancy may degrade the proposed communication model due to the insufficient task-relevant information are encoded into the representation. This validates the tradeoff between the robustness to channel quality variations and the informativeness of the encoded representation.
 
         \begin{figure}
		\centering
		\subfloat[]{
			\centering
			\includegraphics[width=0.47\linewidth]{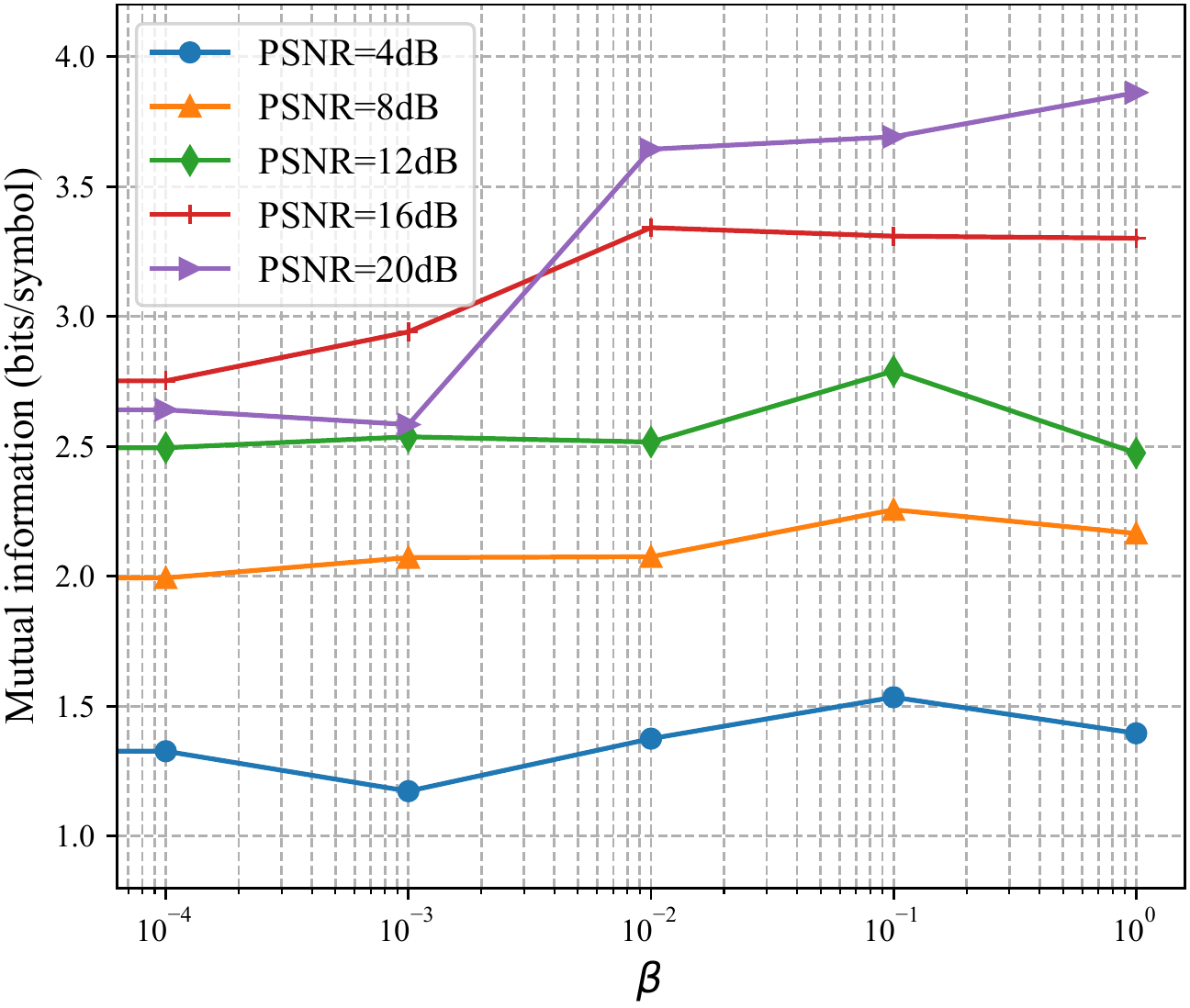}\label{fig:ablation-channelCapacity-0}
		}
		\subfloat[]{
			\centering
			\includegraphics[width=0.47\linewidth]{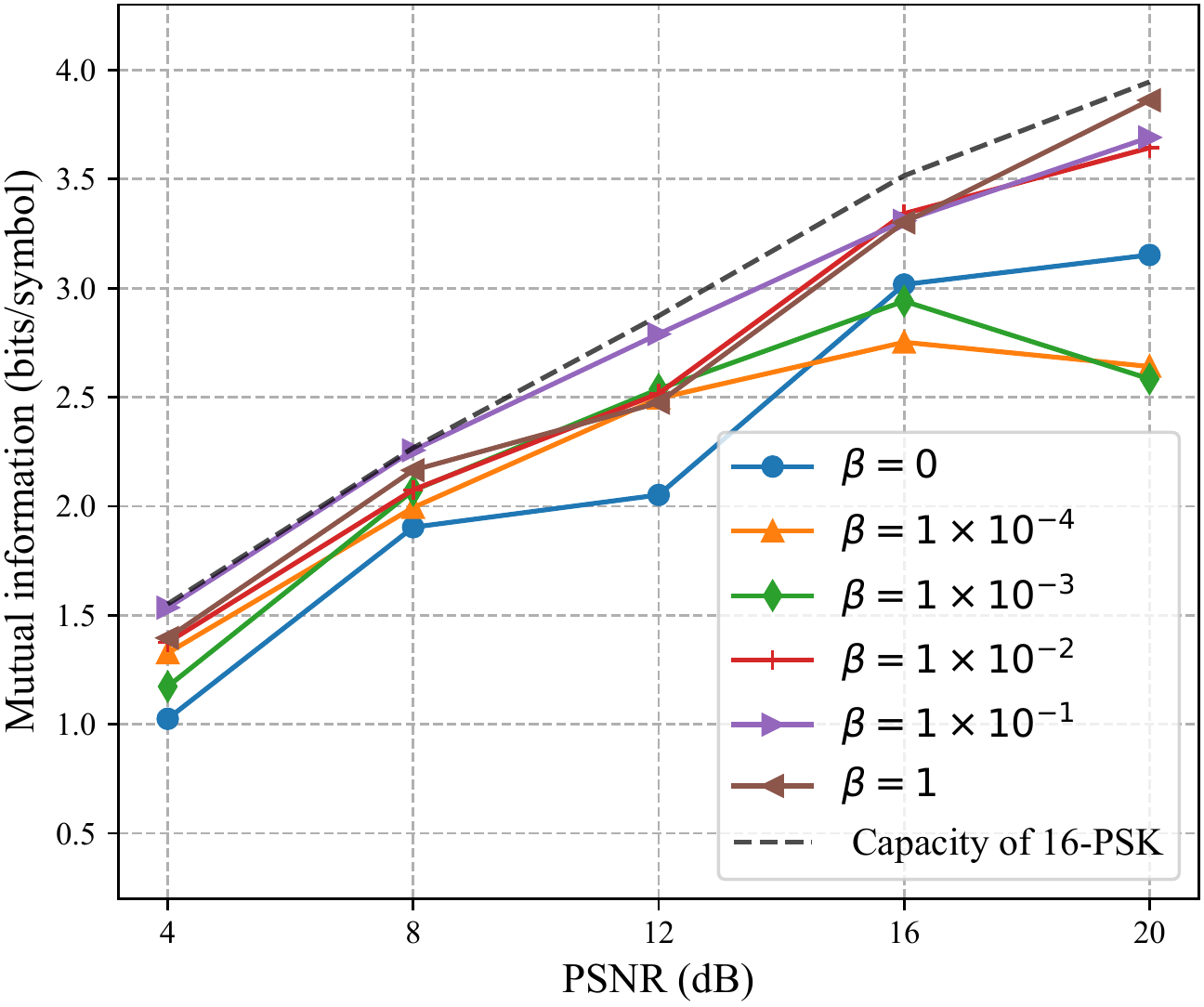}\label{fig:ablation-channelCapacity-1}
		}
		\caption{The mutual information $I(Z;\hZ)$ learned from RIB objectives with different values of $\beta \in [0,1]$ under AWGN channels with varying levels of $\text{PSNR} \in \{4\text{dB}, 8\text{dB}, 12\text{dB}, 16\text{dB}, 20\text{dB} \}$, shown as (a) functions of $\beta$ and (b) functions of $\text{PSNR}$.}
		\label{fig:ablation-channelCapacity}
	\end{figure}
        We conducted experiments to evaluate the effectiveness of the RIB framework in maximizing transmission rate as stated in Remark~\ref{rem:1}, by examining the value of $I(Z; \hZ)$ learned using RIB objectives with different $\beta \in [0,1]$. The experimental results are depicted as functions of $\beta$ and functions of $\text{PSNR}$ in Fig.~\ref{fig:ablation-channelCapacity-0} and Fig.~\ref{fig:ablation-channelCapacity-1}, respectively.
In Fig. \ref{fig:ablation-channelCapacity-0}, we observe that the value of $I(Z; \hZ)$ remains relatively constant over the range of $\beta \in [10^{-4}, 1]$ when the channel conditions are poor (i.e., $\text{PSNR} \in {4\text{dB}, 8\text{dB}, 12\text{dB}}$). In Fig. \ref{fig:ablation-channelCapacity-1}, we find that larger values of $\beta$ (i.e., $\beta \in [10^{-2}, 1]$) lead to $I(Z;\hZ)$ values that are close to the capacity of $16$-PSK modulation. These results demonstrate that minimizing the RIB objectives can lead to an optimal marginal distribution of $\z$ that maximizes the transmission rate, as indicated in Remark~\ref{rem:1}. Furthermore, we observed that smaller values of $\beta$ in good channel conditions result in values of $I(Z; \hZ)$ substantially smaller than the capacity. This is because in this case  the transmitter only transmits signals mostly containing task-relevant information (constrained by the number of classes  $I(Y;\hZ) \leq \log 10$) with little coding redundancy, and needs not to take up channel capacity. Consequently,  when faced with poor channel conditions or a need for prioritizing robustness (e.g., by setting a larger value of $\beta$), the proposed RIB framework facilitates task-oriented communication systems to attain a transmission rate approaching the capacity of extended channels.
 \subsubsection{Feature Splitting and Codebook Size} \label{subsubsec:codebook_size}
 \begin{table*}
		\centering
		\caption{The impact of feature splitting on the number of parameters of codebook $\mathbf{M}$ and the accuracy of DT-JSCC for varying values of $d$ (the number of partitions for feature splitting) and $K$ (the number of codewords).} 
		\label{tab:feature-splitting}
		\resizebox{.95\linewidth}{!}{
			\begin{tabular}{c|cccccccc}
				\toprule
                        & \multicolumn{2}{c}{$K$=8} & \multicolumn{2}{c}{$K$=16} & \multicolumn{2}{c}{$K$=24} & \multicolumn{2}{c}{$K$=32}\\
				$d$ & \# parameters & Accuracy & \# parameters & Accuracy & \# parameters & Accuracy & \# parameters & Accuracy\\
				\midrule
				1 & 8192 & 42.21& 16384& 46.97& 24576 & 50.49 & 32768 & 51.59\\
                    2 & 4092 & 78.35& 8192& 78.97& 12288 & 78.28 &16384 & 63.87\\
                    4 & 2048 & 89.60& 4096& 92.74& 6144 & 94.99 & 8192 & 93.52 \\
                    8 & 1024 & 95.24& 2048& 95.62& 3072 & 96.04 & 4096 & 96.05 \\
                    16& \underline{512} & \underline{97.13}& \underline{1024}& \underline{97.27}& \underline{1536} & \underline{97.15} & \underline{2048} & \underline{97.07} \\
                    32& \textbf{256} & \textbf{97.78}& \textbf{512} & \textbf{97.64}& \textbf{768} & \textbf{97.41} & \textbf{1024} & \textbf{97.57} \\
				\bottomrule
		\end{tabular}}
	\end{table*}
	\begin{figure}
		\centering
		\includegraphics[width=8.8cm]{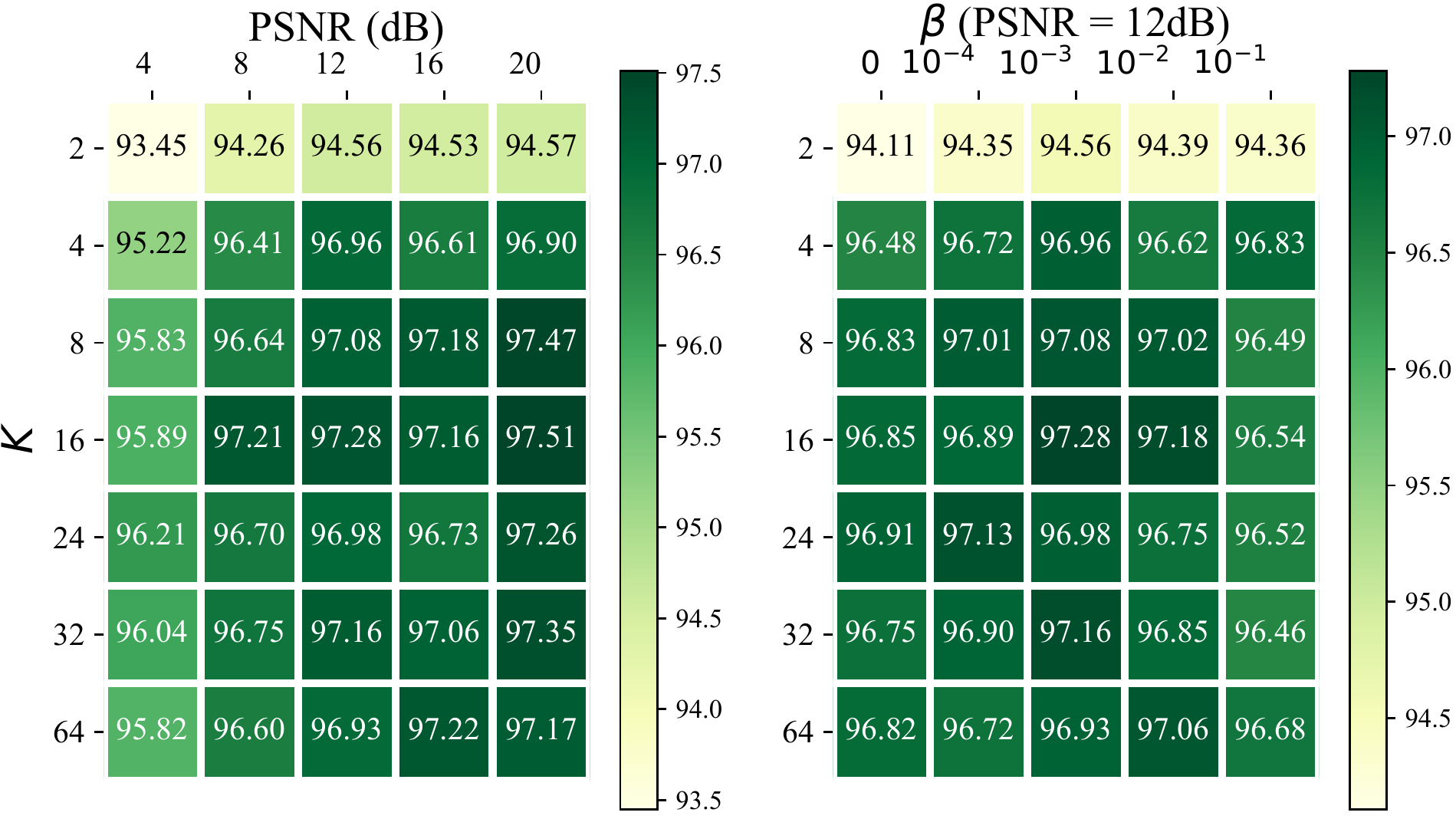}
		\caption{Performance comparison of DT-JSCC with different codebook sizes on MNIST classification task over AWGN channels with different PSNR and different values of $\beta$.} 
		\label{fig:ablation-codebooksize}
	\end{figure}
In this section, we examine the impact of feature splitting and codebook size through the ablation experiments on the MINST classification tasks. Since the robustness of RIB framework has been confirmed in previous ablation studies, we conduct the following experiments by training and evaluating the systems under AWGN channel models with the same PSNR.
 
We first conduct experiments by varying the number of partitions for feature splitting $d$ and the number of codewords $K$. It is worth noting that feature splitting can affect the size of codebook $\mathbf{M}$ due to the dimension of split vectors being that of codewords. To assess its impact on memory occupation of $\mathbf{M}$ and the inference performance, we present the number of parameters of $\mathbf{M}$ and the accuracy of DT-JSCC in Table \ref{tab:feature-splitting}. It is evident that splitting features into more partitions can significantly decrease the parameters of $\mathbf{M}$ and enhance the inference performance. However, the number of partitions for feature splitting $d$ is also the dimension of $\z$ that determines the communication latency. Therefore, it is essential to choose an appropriate feature splitting strategy that balances usage, inference performance, and communication latency.
 
Then, we investigate the impact of codebook size $K$ on the inference performance of DT-JSCC and its correlation with the value of $\beta$ in the RIB framework. Initially, we fix the hyperparameter $\beta = 0.001$ and vary the codebook size $K \in [2, 64]$ over AWGN channels with different PSNR. Subsequently, we vary the codebook size $K \in [2, 64]$ with different $\beta$ over AWGN channels with $\text{PSNR} = 12 \text{dB}$.
 The experimental results depicted in Fig. \ref{fig:ablation-codebooksize} reveal that  increasing the codebook size $K$ leads to a gradual improvement in inference performance, but the performance soon saturates. Moreover, DT-JSCC systems with large codebooks perform similarly to those with an appropriate codebook size of around $16$. 
 It is worth noting that the capacity of the extended channel is determined by $K$, and is also the upper limit of task-relevant information and redundancy controlled by $\beta$, as stated in our theoretical analysis (refer to Figure \ref{fig:illu_tradeoff}). 
 While increasing $K$ can increase the upper bound, the task-relevant information is limited by the cardinality of the target variable and cannot be increased. Additionally, the experimental results show that the DT-JSCC systems trained by RIB with different $\beta$ achieve similar inference accuracy when training channels and evaluating channels have the same PSNR. This verifies our theoretical insight that increasing the number of codewords allows more redundancy in the communication model (i.e., allows for a larger value of $\beta$), but cannot improve the inference performance, indicating that  there exists no direct correlation between the codebook size
and the value of $\beta$ when the channel quality remains the same. 

\section{Discussion}
	\label{sec:dis}
	In this work, the RIB principle is proposed to include the coded redundancy for robust channel encoding, and the neural vector quantization is proposed in the task-oriented communication system to learn the compact discrete representation. 
	The above experiments empirically demonstrate the superiority of our proposed DT-JSCC model on the aspects of inference performance and the robustness to channel quality fluctuation. However, there are two primary concerns that we identify in this work. The first concern pertains to the communication overhead incurred by the coded redundancy, while the second concerns the memory cost associated with the learnable codebook residing on both the transmitter and receiver. We discuss each of these concerns in detail below.
	\begin{itemize}
		\item The communication overhead is determined by the dimensions over which the channel symbols are transmitted.  For the task-oriented communication system with continuous representation, the encoded redundancy may expand the dimensionality of the encoded representation. The recent work proposes to reduce the communication overhead by pruning the redundant dimensions of the encoded continuous representation. However, for the discrete task-oriented communication system, DT-JSCC, the encoded redundancy is not necessarily induced by the redundant dimensions. A large codebook with redundant codewords can provide sufficient redundancy in the discrete representation. Therefore, the proposed DT-JSCC method can encode additional redundancy without increasing the communication overhead, but at the expense of expanding the storage space for a large codebook.
		\item From the ablation experiments in Section~\ref{subsec:ablation}, we can observe that the performance of the system model with large codebooks is similar to that of the model with an appropriate codebook size. Theoretically, the more complex inference task requires more task-relevant information transmitted to the receiver so that the larger codebooks consume more memory on the transmitter and the receiver. After all, there exists an inherent tradeoff between the informativeness of the discrete representation and the memory resource saving on the communication system. 
	\end{itemize}
	\section{Conclusion}
	\label{sec:con}
        In this work, we investigate the tradeoff between the informativeness of the encoded representation and the robustness against channel variation in task-oriented communications. To formalize the informativeness-robustness tradeoff, we propose a theoretical framework, named robust information bottleneck (RIB). Our proposed framework enhances the robustness of the communication systems by utilizing the coded redundancy in the encoded representation. Since the computation intractability of the mutual information, we adopt a variational upper bound of the objective function by a variational distribution parameterized by DNNs. 
        
        Furthermore, we propose a task-oriented communication scheme with digital modulation, DT-JSCC, where informative messages for inference tasks are encoded into discrete representations. It is compatible with canonical finite-point constellations in modern mobile systems. Our simulation results demonstrate that the proposed method outperforms the baselines with low communication latency. With the above advantages of DT-JSCC, one can further explore the usage of the proposed method for other AI applications in 6G, e.g., the deployment of generative models on the network edge, reliable semantic information transmission, privacy-sensitive task-oriented communication, etc.

\bibliographystyle{IEEEtran}
\bibliography{IEEEabrv,ref}          

\begin{thebibliography}{10}
\providecommand{\url}[1]{#1}
\csname url@samestyle\endcsname
\providecommand{\newblock}{\relax}
\providecommand{\bibinfo}[2]{#2}
\providecommand{\BIBentrySTDinterwordspacing}{\spaceskip=0pt\relax}
\providecommand{\BIBentryALTinterwordstretchfactor}{4}
\providecommand{\BIBentryALTinterwordspacing}{\spaceskip=\fontdimen2\font plus
\BIBentryALTinterwordstretchfactor\fontdimen3\font minus
  \fontdimen4\font\relax}
\providecommand{\BIBforeignlanguage}[2]{{%
\expandafter\ifx\csname l@#1\endcsname\relax
\typeout{** WARNING: IEEEtran.bst: No hyphenation pattern has been}%
\typeout{** loaded for the language `#1'. Using the pattern for}%
\typeout{** the default language instead.}%
\else
\language=\csname l@#1\endcsname
\fi
#2}}
\providecommand{\BIBdecl}{\relax}
\BIBdecl

\bibitem{anthes2016state}
C.~Anthes, R.~J. Garc{\'\i}a-Hern{\'a}ndez, M.~Wiedemann, and
  D.~Kranzlm{\"u}ller, ``State of the art of virtual reality technology,'' in
  \emph{2016 IEEE aerospace conference}.\hskip 1em plus 0.5em minus 0.4em\relax
  IEEE, 2016, pp. 1--19.

\bibitem{voulodimos2018deep}
A.~Voulodimos, N.~Doulamis, A.~Doulamis, and E.~Protopapadakis, ``Deep learning
  for computer vision: A brief review,'' \emph{Computational intelligence and
  neuroscience}, vol. 2018, 2018.

\bibitem{van2012brain}
J.~Van~Erp, F.~Lotte, and M.~Tangermann, ``Brain-computer interfaces: beyond
  medical applications,'' \emph{Computer}, vol.~45, no.~4, pp. 26--34, 2012.

\bibitem{ye2018machine}
H.~Ye, L.~Liang, G.~Y. Li, J.~Kim, L.~Lu, and M.~Wu, ``Machine learning for
  vehicular networks: Recent advances and application examples,'' \emph{ieee
  vehicular technology magazine}, vol.~13, no.~2, pp. 94--101, 2018.

\bibitem{zhou2021review}
S.~K. Zhou, H.~Greenspan, C.~Davatzikos, J.~S. Duncan, B.~Van~Ginneken,
  A.~Madabhushi, J.~L. Prince, D.~Rueckert, and R.~M. Summers, ``A review of
  deep learning in medical imaging: Imaging traits, technology trends, case
  studies with progress highlights, and future promises,'' \emph{Proceedings of
  the IEEE}, vol. 109, no.~5, pp. 820--838, 2021.

\bibitem{letaief2019roadmap}
K.~B. Letaief, W.~Chen, Y.~Shi, J.~Zhang, and Y.-J.~A. Zhang, ``The roadmap to
  6g: Ai empowered wireless networks,'' \emph{IEEE communications magazine},
  vol.~57, no.~8, pp. 84--90, 2019.

\bibitem{strinati20216g}
E.~C. Strinati and S.~Barbarossa, ``6g networks: Beyond shannon towards
  semantic and goal-oriented communications,'' \emph{Computer Networks}, vol.
  190, p. 107930, 2021.

\bibitem{hoydis2021toward}
J.~Hoydis, F.~A. Aoudia, A.~Valcarce, and H.~Viswanathan, ``Toward a 6g
  ai-native air interface,'' \emph{IEEE Communications Magazine}, vol.~59,
  no.~5, pp. 76--81, 2021.

\bibitem{shao2021learning}
J.~Shao, Y.~Mao, and J.~Zhang, ``Learning task-oriented communication for edge
  inference: An information bottleneck approach,'' \emph{IEEE Journal on
  Selected Areas in Communications}, vol.~40, no.~1, pp. 197--211, 2021.

\bibitem{letaief2021edge}
K.~B. Letaief, Y.~Shi, J.~Lu, and J.~Lu, ``Edge artificial intelligence for 6g:
  Vision, enabling technologies, and applications,'' \emph{IEEE Journal on
  Selected Areas in Communications}, vol.~40, no.~1, pp. 5--36, 2021.

\bibitem{shi2023taskoriented}
Y.~Shi, Y.~Zhou, D.~Wen, Y.~Wu, C.~Jiang, and K.~B. Letaief, ``Task-oriented
  communications for 6g: Vision, principles, and technologies,'' \emph{arXiv
  preprint arXiv:2303.10920}, 2023.

\bibitem{bourtsoulatze2019deep}
E.~Bourtsoulatze, D.~B. Kurka, and D.~G{\"u}nd{\"u}z, ``Deep joint
  source-channel coding for wireless image transmission,'' \emph{IEEE
  Transactions on Cognitive Communications and Networking}, vol.~5, no.~3, pp.
  567--579, 2019.

\bibitem{choi2019neural}
K.~Choi, K.~Tatwawadi, A.~Grover, T.~Weissman, and S.~Ermon, ``Neural joint
  source-channel coding,'' in \emph{International Conference on Machine
  Learning}.\hskip 1em plus 0.5em minus 0.4em\relax PMLR, 2019, pp. 1182--1192.

\bibitem{guyader2001joint}
A.~Guyader, E.~Fabre, C.~Guillemot, and M.~Robert, ``Joint source-channel turbo
  decoding of entropy-coded sources,'' \emph{IEEE Journal on Selected Areas in
  Communications}, vol.~19, no.~9, pp. 1680--1696, 2001.

\bibitem{wei2021low}
K.~Wei, J.~Li, C.~Ma, M.~Ding, C.~Chen, S.~Jin, Z.~Han, and H.~V. Poor,
  ``Low-latency federated learning over wireless channels with differential
  privacy,'' \emph{IEEE Journal on Selected Areas in Communications}, vol.~40,
  no.~1, pp. 290--307, 2021.

\bibitem{shafi20175g}
M.~Shafi, A.~F. Molisch, P.~J. Smith, T.~Haustein, P.~Zhu, P.~De~Silva,
  F.~Tufvesson, A.~Benjebbour, and G.~Wunder, ``5g: A tutorial overview of
  standards, trials, challenges, deployment, and practice,'' \emph{IEEE journal
  on selected areas in communications}, vol.~35, no.~6, pp. 1201--1221, 2017.

\bibitem{bhagyaveni2016introduction}
M.~Bhagyaveni, R.~Kalidoss, and K.~Vishvaksenan, \emph{Introduction to analog
  and digital communication}.\hskip 1em plus 0.5em minus 0.4em\relax River
  Publishers, 2016, vol.~46.

\bibitem{shao2020bottlenet++}
J.~Shao and J.~Zhang, ``Bottlenet++: An end-to-end approach for feature
  compression in device-edge co-inference systems,'' in \emph{2020 IEEE
  International Conference on Communications Workshops (ICC Workshops)}.\hskip
  1em plus 0.5em minus 0.4em\relax IEEE, 2020, pp. 1--6.

\bibitem{tishby2000information}
N.~Tishby, F.~C. Pereira, and W.~Bialek, ``The information bottleneck method,''
  \emph{arXiv preprint physics/0004057}, 2000.

\bibitem{shao2022task}
J.~Shao, Y.~Mao, and J.~Zhang, ``Task-oriented communication for multi-device
  cooperative edge inference,'' \emph{IEEE Transactions on Wireless
  Communications}, 2022.

\bibitem{aguerri2019distributed}
I.~E. Aguerri and A.~Zaidi, ``Distributed variational representation
  learning,'' \emph{IEEE transactions on pattern analysis and machine
  intelligence}, vol.~43, no.~1, pp. 120--138, 2019.

\bibitem{xie2021deep}
H.~Xie, Z.~Qin, G.~Y. Li, and B.-H. Juang, ``Deep learning enabled semantic
  communication systems,'' \emph{IEEE Transactions on Signal Processing},
  vol.~69, pp. 2663--2675, 2021.

\bibitem{jankowski2020wireless}
M.~Jankowski, D.~G{\"u}nd{\"u}z, and K.~Mikolajczyk, ``Wireless image retrieval
  at the edge,'' \emph{IEEE Journal on Selected Areas in Communications},
  vol.~39, no.~1, pp. 89--100, 2020.

\bibitem{shannon1948mathematical}
C.~E. Shannon, ``A mathematical theory of communication,'' \emph{The Bell
  system technical journal}, vol.~27, no.~3, pp. 379--423, 1948.

\bibitem{goldsmith1995joint}
A.~Goldsmith, ``Joint source/channel coding for wireless channels,'' in
  \emph{1995 IEEE 45th Vehicular Technology Conference. Countdown to the
  Wireless Twenty-First Century}, vol.~2.\hskip 1em plus 0.5em minus
  0.4em\relax IEEE, 1995, pp. 614--618.

\bibitem{zhai2005joint}
F.~Zhai, Y.~Eisenberg, and A.~K. Katsaggelos, ``Joint source-channel coding for
  video communications,'' \emph{Handbook of Image and Video Processing}, pp.
  1065--1082, 2005.

\bibitem{fresia2010joint}
M.~Fresia, F.~Perez-Cruz, H.~V. Poor, and S.~Verdu, ``Joint source and channel
  coding,'' \emph{IEEE Signal Processing Magazine}, vol.~27, no.~6, pp.
  104--113, 2010.

\bibitem{chen2018joint}
C.~Chen, L.~Wang, and F.~C. Lau, ``Joint optimization of protograph ldpc code
  pair for joint source and channel coding,'' \emph{IEEE Transactions on
  Communications}, vol.~66, no.~8, pp. 3255--3267, 2018.

\bibitem{kurka2020deepjscc}
D.~B. Kurka and D.~G{\"u}nd{\"u}z, ``Deepjscc-f: Deep joint source-channel
  coding of images with feedback,'' \emph{IEEE Journal on Selected Areas in
  Information Theory}, vol.~1, no.~1, pp. 178--193, 2020.

\bibitem{farsad2018deep}
N.~Farsad, M.~Rao, and A.~Goldsmith, ``Deep learning for joint source-channel
  coding of text,'' in \emph{2018 IEEE international conference on acoustics,
  speech and signal processing (ICASSP)}.\hskip 1em plus 0.5em minus
  0.4em\relax IEEE, 2018, pp. 2326--2330.

\bibitem{saidutta2021joint}
Y.~M. Saidutta, A.~Abdi, and F.~Fekri, ``Joint source-channel coding over
  additive noise analog channels using mixture of variational autoencoders,''
  \emph{IEEE Journal on Selected Areas in Communications}, vol.~39, no.~7, pp.
  2000--2013, 2021.

\bibitem{dai2022nonlinear}
J.~Dai, S.~Wang, K.~Tan, Z.~Si, X.~Qin, K.~Niu, and P.~Zhang, ``Nonlinear
  transform source-channel coding for semantic communications,'' \emph{IEEE
  Journal on Selected Areas in Communications}, 2022.

\bibitem{wang2022constellation}
M.~Wang, J.~Li, M.~Ma, and X.~Fan, ``Constellation design for deep joint
  source-channel coding,'' \emph{IEEE Signal Processing Letters}, vol.~29, pp.
  1442--1446, 2022.

\bibitem{xie2020lite}
H.~Xie and Z.~Qin, ``A lite distributed semantic communication system for
  internet of things,'' \emph{IEEE Journal on Selected Areas in
  Communications}, vol.~39, no.~1, pp. 142--153, 2020.

\bibitem{zhu2019joint}
B.~Zhu, J.~Wang, L.~He, and J.~Song, ``Joint transceiver optimization for
  wireless communication phy using neural network,'' \emph{IEEE Journal on
  Selected Areas in Communications}, vol.~37, no.~6, pp. 1364--1373, 2019.

\bibitem{blei2017variational}
D.~M. Blei, A.~Kucukelbir, and J.~D. McAuliffe, ``Variational inference: A
  review for statisticians,'' \emph{Journal of the American statistical
  Association}, vol. 112, no. 518, pp. 859--877, 2017.

\bibitem{kullback1951information}
S.~Kullback and R.~A. Leibler, ``On information and sufficiency,'' \emph{The
  annals of mathematical statistics}, vol.~22, no.~1, pp. 79--86, 1951.

\bibitem{cover1999elements}
T.~M. Cover, \emph{Elements of information theory}.\hskip 1em plus 0.5em minus
  0.4em\relax John Wiley \& Sons, 1999.

\bibitem{kingma2013auto}
D.~P. Kingma and M.~Welling, ``Auto-encoding variational bayes,'' in \emph{2nd
  International Conference on Learning Representations, {ICLR} 2014, Banff, AB,
  Canada, April 14-16, 2014, Conference Track Proceedings}, 2014.

\bibitem{alemi2016deep}
A.~A. Alemi, I.~Fischer, J.~V. Dillon, and K.~Murphy, ``Deep variational
  information bottleneck,'' in \emph{5th International Conference on Learning
  Representations, {ICLR} 2017, Toulon, France, April 24-26, 2017, Conference
  Track Proceedings}, 2017.

\bibitem{jang2016categorical}
E.~Jang, S.~Gu, and B.~Poole, ``Categorical reparameterization with
  gumbel-softmax,'' in \emph{5th International Conference on Learning
  Representations, {ICLR} 2017, Toulon, France, April 24-26, 2017, Conference
  Track Proceedings}, 2017.

\bibitem{baevski2020wav2vec}
A.~Baevski, Y.~Zhou, A.~Mohamed, and M.~Auli, ``wav2vec 2.0: A framework for
  self-supervised learning of speech representations,'' \emph{Advances in
  Neural Information Processing Systems}, vol.~33, pp. 12\,449--12\,460, 2020.

\bibitem{van2017neural}
A.~Van Den~Oord, O.~Vinyals \emph{et~al.}, ``Neural discrete representation
  learning,'' \emph{Advances in neural information processing systems},
  vol.~30, 2017.

\bibitem{razavi2019generating}
A.~Razavi, A.~Van~den Oord, and O.~Vinyals, ``Generating diverse high-fidelity
  images with vq-vae-2,'' \emph{Advances in neural information processing
  systems}, vol.~32, 2019.

\bibitem{tao2018digital}
F.~Tao, H.~Zhang, A.~Liu, and A.~Y. Nee, ``Digital twin in industry:
  State-of-the-art,'' \emph{IEEE Transactions on industrial informatics},
  vol.~15, no.~4, pp. 2405--2415, 2018.

\bibitem{maier2020internet}
M.~Maier, A.~Ebrahimzadeh, S.~Rostami, and A.~Beniiche, ``The internet of no
  things: Making the internet disappear and" see the invisible",'' \emph{IEEE
  Communications Magazine}, vol.~58, no.~11, pp. 76--82, 2020.

\bibitem{lecun1998gradient}
Y.~LeCun, L.~Bottou, Y.~Bengio, and P.~Haffner, ``Gradient-based learning
  applied to document recognition,'' \emph{Proceedings of the IEEE}, vol.~86,
  no.~11, pp. 2278--2324, 1998.

\bibitem{krizhevsky2009learning}
A.~Krizhevsky, ``Learning multiple layers of features from tiny images,''
  University of Toronto, Tech. Rep., 2009.

\bibitem{kingma2015variational}
D.~P. Kingma, T.~Salimans, and M.~Welling, ``Variational dropout and the local
  reparameterization trick,'' \emph{Advances in neural information processing
  systems}, vol.~28, 2015.

\bibitem{he2016deep}
K.~He, X.~Zhang, S.~Ren, and J.~Sun, ``Deep residual learning for image
  recognition,'' in \emph{Proceedings of the IEEE conference on computer vision
  and pattern recognition}, 2016, pp. 770--778.

\end{thebibliography}

\end{document}